\documentclass[twocolumn,pre]{revtex4-1}
\usepackage{color}
\usepackage{graphics,graphicx,epsfig,wrapfig}

\usepackage{epsf,epstopdf}
\usepackage{amssymb,amsfonts,amsmath}
\usepackage{enumitem}
\usepackage{setspace}

\usepackage{ifthen}
\usepackage{placeins} 
\usepackage{siunitx}

\newcommand\mydots{\hbox to 0.8em{.\hss.\hss.}}
\newcommand{\beqn}{\begin{eqnarray}}
\newcommand{\eeqn}{\end{eqnarray}}
\newcommand{\beq}{\begin{equation}}
\newcommand{\eeq}{\end{equation}}

\newcommand{\ch}{}

\newcommand{\ff}{1}

\newcommand{\mytitle}{Primary and secondary anti-viral response captured by the dynamics and phenotype of individual T cell clones}
\newcommand{\myauthors}{Anastasia A. Minervina$^{1}$, Mikhail V. Pogorelyy$^{1,2}$, Ekaterina A. Komech$^{1,2}$,
Vadim K. Karnaukhov$^{3}$, Petra Bacher$^{4,5}$, Elisa Rosati$^{5}$, Andre Franke$^{5}$, Dmitriy M. Chudakov$^{1,2,3,6}$, Ilgar Z. Mamedov$^{1,6,7}$, \\ Yuri B. Lebedev*$^{1,8}$, Thierry Mora*$^{9}$, Aleksandra M. Walczak*$^{9}$}

\makeatletter
\def\@seccntformat#1{%
  \expandafter\ifx\csname c@#1\endcsname\c@section\else
  \csname the#1\endcsname\quad
  \fi}
\makeatother

\begin{document}
\title{\mytitle}
\author{\myauthors}
\affiliation{~\\
\normalsize{$^{1}$ Shemyakin-Ovchinnikov Institute of Bioorganic
  Chemistry,}
\normalsize{Moscow, Russia}\\
\normalsize{$^{2}$Pirogov Russian National Research Medical University, Moscow, Russia}\\
\normalsize{$^{3}$Center of Life Sciences, Skoltech, Moscow, Russia}\\
\normalsize{$^{4}$Institute of Immunology, Kiel University, Kiel, Germany}\\
\normalsize{$^{5}$Institute of Clinical Molecular Biology, Kiel University, Kiel, Germany}\\
\normalsize{$^{6}$Masaryk University, Central European Institute of Technology, Brno, Czech Republic}\\
\normalsize{$^{7}$V.I. Kulakov National Medical Research Center for Obstetrics, Gynecology and Perinatology, Moscow, Russia}\\
 \normalsize{$^{8}$Moscow State University, Moscow, Russia}\\
\normalsize{$^{9}$Laboratoire de physique de l'\'Ecole normale sup\'erieure,}
\normalsize{ENS, PSL, Sorbonne Universit\'e, Universit\'e de Paris, and CNRS, 75005 Paris,
  France}\\
\normalsize{\rm *These authors contributed equally.}\\
}

\begin{abstract} 
{
The diverse repertoire of T-cell receptors (TCR) plays a key role in the adaptive immune response to infections. 
Using TCR alpha and beta repertoire sequencing for T-cell subsets, as well as single-cell RNAseq and TCRseq, we track the concentrations and phenotypes of individual T-cell clones in response to primary and secondary yellow fever immunization --- the model for acute infection in humans --- showing their large diversity. We confirm the secondary response is an order of magnitude weaker, albeit $\sim10$ days faster than the primary one. Estimating the fraction of the T-cell response directed against the single immunodominant epitope, we identify the sequence features of TCRs that define the high precursor frequency of the two major TCR motifs specific for this particular epitope. We also show the consistency of clonal expansion dynamics between bulk alpha and beta repertoires, using a new  methodology to reconstruct alpha-beta pairings from clonal trajectories. 
}
\end{abstract}

\maketitle

\section*{Introduction}
T-cells play a crucial role in the immune response to pathogens by mediating antibody formation and clearance of infected cells, and by defining an overall response strategy. The specificity of T-cells is determined by the T-cell receptor (TCR), a heterodimer of alpha and beta protein chains. Genes for alpha and beta chains assemble in a random process of somatic V(D)J-recombination, which leads to a huge variety of possible TCRs \citep{murugan_statistical_2012}. The resulting diverse na\"ive repertoire contains T-cell clones that recognize epitopes of yet unseen pathogens, and can participate in the immune response to infection or vaccination. 
One of the best established models of acute viral infection in humans is yellow fever (YF) vaccination. Yellow fever vaccine is a live attenuated virus with a peak of viremia happening around day 7 after vaccine administration \citep{miller_human_2008,akondy_yellow_2009,akondy_initial_2015}. The dynamics of primary T-cell response was investigated by various techniques: cell activation marker staining \citep{miller_human_2008,blom_temporal_2013,kohler_early_2012,kongsgaard_adaptive_2017}, MHC multimer staining \citep{akondy_yellow_2009,blom_temporal_2013,james_yellow_2013,kongsgaard_adaptive_2017}, high-throughput sequencing \citep{dewitt_dynamics_2015,pogorelyy_precise_2018} and deuterium cell labelling \citep{akondy_origin_2017}. Primary T-cell response sharply peaks around 2 weeks after YFV17D (vaccine strain of yellow fever virus) vaccination \citep{miller_human_2008,akondy_yellow_2009,kohler_early_2012,pogorelyy_precise_2018,james_yellow_2013}. The immune response is very diverse and targets multiple epitopes inside the YF virus \citep{de_melo_t-cell_2013,co_human_2002,akondy_yellow_2009,james_yellow_2013,blom_temporal_2013}. 
An essential feature of effective vaccination is the formation of immune memory. Although most of the effector cells die shortly after viral clearance, YF-specific T-cells could be found in the blood of vaccinated individuals years \citep{akondy_yellow_2009,akondy_origin_2017,kongsgaard_adaptive_2017,james_yellow_2013} and even decades after vaccination \citep{fuertes_marraco_long-lasting_2015,wieten_single_2016}. While the immune response to the primary vaccination has been much studied, there is only limited data on the response to the booster vaccination with YFV17D. Both T-cell activation marker staining and multimer staining show that the secondary response is much weaker than the primary one \citep{kongsgaard_adaptive_2017}, but their precise dynamics, diversity, and clonal structure are still unknown. 

In summary, previous studies provide insight into the macroscopic features of the T-cell response, such as total frequency of T-cells with an activated phenotype, or T-cells specific to a particular viral epitope on different timepoints after vaccination. However, with recently developed methods it is now possible to uncover the microscopic structure of the primary and secondary immune response, such as the dynamics and phenotypes of distinct T-cell clones, as well as the receptor features that determine the recognition of epitopes.

TCR repertoire sequencing allows for longitudinally tracking individual clones of responding T-cells irrespective of their epitope specificity. Single-cell RNAseq (scRNAseq) enables simultaneous quantification of thousands of transcripts per cell for thousands of cells, providing an unbiased characterization of immune cell phenotype. Single-cell TCR sequencing produces paired $\alpha\beta$ repertoire data, and thus could help discover conserved sequence motifs in one or both TCR chains. These motifs encode TCR structural features essential to antigen recognition \citep{dash_quantifiable_2017,glanville_identifying_2017}. Information about complete TCR sequences allows homological modeling of TCR structure \citep{schritt_repertoire_2019}, which can be used for binding prediction with protein-protein docking \citep{pierce_flexible_2013}.
We combine longitudinal TCR alpha and beta repertoire sequencing, scRNAseq, scTCRseq, TCR structure modelling and TCR-pMHC docking simulations to get a comprehensive picture of primary and secondary T-cell response to the yellow fever vaccine -- the \textit{in vivo} model of acute viral infection in humans.

\section*{Results}

\begin{figure*}
\noindent\includegraphics[width=\ff\linewidth]{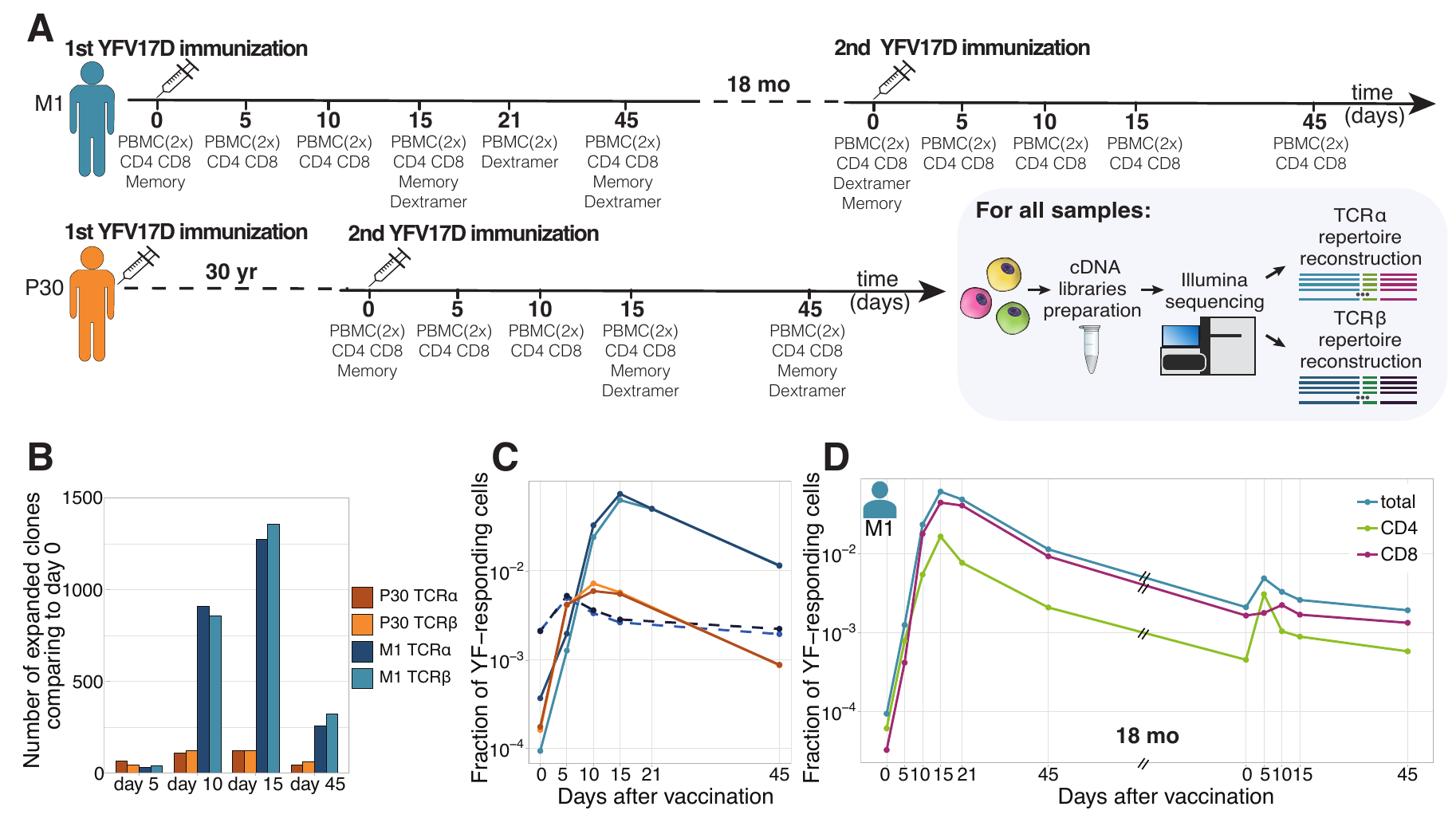}
\caption{{\bf Primary and secondary response to yellow fever vaccination. A.} Experiment design. Blood was taken at multiple timepoints before and after primary and secondary immunization against yellow fever virus. Two biological replicates of PBMCs and different cell subpopulations (indicated below each day of blood draw) were isolated at all timepoints. cDNA TCR alpha and TCR beta libraries were sequenced on Illumina platform. {\bf B.} The number of significantly expanded TCR alpha and TCR beta clonotypes for both donors in comparison to day 0. For donor P30 the number of significantly expanded clones is lower, than observed in primary vaccinations (see Fig. S2). {\bf C.} {\ch The fraction of YF-responding cells as a proportion of all T-cells, measured by} cumulative frequency of {\ch YF-responding} TCR alpha and beta clonotypes of donor M1 after first (light blue and dark blue) and second immunization (dashed light blue and dark blue), and donor P30 (orange and yellow), which had a second immunization 30 years after the first. {\bf D.} {\ch The fraction of} CD4+ and CD8+ YF-responding {\ch cells, as a proportion of all T-cells} of donor M1 during the primary and secondary response to YFV17D. {\ch No novel major expansions were observed after secondary immunization, see Fig. S1}
}
\label{fig1}
\end{figure*}

\subsection*{Secondary T-cell response to the YFV17D vaccine is weaker but faster than the primary response}
We sequenced TCR alpha and TCR beta repertoires of bulk peripheral blood mononuclear cells (PBMCs) and different T-cell subsets at multiple timepoints before and after primary and booster vaccination against yellow fever of donor M1 (Fig. 1A). Clonotypes responding to the primary YF immunization were identified using the edgeR software as previously described \citep{pogorelyy_precise_2018}. Briefly, the biological replicates of bulk PBMCs were used to estimate the noise in the TCR mRNA counts. Clonotypes were assumed YF-responding if they increased in concentration more than 32-fold ($p<0.01$, {\ch see Methods})  between any two timepoints before the peak of the primary response (days 0, 5, 10 and 15).

Overall we found 1580 TCR beta and 1566 TCR alpha clonotypes significantly expanded after the primary immunization, respectively occupying 6.7\% and 7.8\% of the sampled TCR repertoire {\ch of bulk PBMCs} in cumulative frequency at the peak of the response (Fig. 1B, C). As expected, both the numbers of responding clones and their cumulative frequencies were very similar for expanded clonotypes identified in bulk TCR alpha and beta repertoires. For simplicity in the following sections we focus on TCR beta repertoires, unless stated otherwise. In accordance with previous studies \citep{miller_human_2008,blom_temporal_2013,akondy_yellow_2009,kongsgaard_adaptive_2017,pogorelyy_precise_2018}, we show that during the primary response T-cells expanded intensely (with cumulative increase of about 950-fold) within 2-3 weeks after  YF immunization. They subsequently contracted, but still exceeded baseline frequency 18 months afterwards.

We then tracked these YF-responding clonotypes identified during primary immunization before and after the second vaccination 18 months after the first one. The cumulative frequency of these clonotypes increased $\approx$2.5-fold at the peak of the response after the second immunization, reaching 0.5\% of the TCR repertoire (Fig. 1D, blue curve). The secondary response was weaker, but happened much faster than the primary one, with a peak frequency of responding clonotypes occurring on day 5 instead of day 15 after vaccination. {\ch To check if there was also recruitment of new clonotypes in the secondary response, we applied edgeR to timepoints from the second immunization only. Although we identified 73 additional responding clonotypes, their impact on the magnitude of the secondary response was negligible and we did not use them for further analyses (see Fig.~S1). Backtracking of these novel clonotypes showed that they also slightly  expanded during the primary response but not enough enough to pass our significance and magnitude thresholds. In summary, we found no evidence of substantial recruitment of naive clones in the response to the booster vaccination.}

Using sequenced CD4+ and CD8+ T-cell subsets, we attributed a CD4 or CD8 phenotype to each responding clone (see Methods) and thus could track these two subsets separately. 
After booster immunization in donor M1, YF-responding CD4+ cells peaked earlier  (day 5 vs day 10) and expanded much more ($\approx8$ times vs. $\approx1.5$ times)  than  CD8+ T-cells (Fig. 1D, green and pink curves). During primary immunization, the difference in response dynamics between CD4+ and CD8+ subsets was less prominent, as they both peaked on day 15. However, by day 21 CD4+ responding clones contracted much more (to 43.6\% of peak frequency) than CD8+ clonotypes (87\% of peak frequency). These observations confirm previous reports that the CD4 response precedes the CD8 response \citep{blom_temporal_2013}. 

\subsection*{Secondary response to booster vaccination after 18 months and after 30 years have similar features}
To see how long-lived T-cell memory response to YF can be, we recruited an additional donor (P30), who received the first YF-vaccine 30 years earlier and has not been in YF endemic areas for at least 28 years. From this donor, we collected bulk PBMCs and several T-cell subsets before and after booster immunization. Both the numbers of responding clonotypes (204 for TCR beta and 201 for TCR alpha) and the maximum frequency at the peak of the response (0.69\%) were much lower than for any primary vaccinee both from this and other studies (Fig. S2). Most of these clonotypes were low frequency or undetected before the second immunization, although a few were sampled in the memory repertoire prior to vaccination.

The response to the booster vaccination was characterized by a large expansion between days 0 and 5, and a peak on day 10, for both CD4+ and CD8+ T-cells.
Overall the dynamics and the magnitude of this response was very similar to the response to the booster vaccination after 18 months we observed in donor M1 (Fig. 1C), suggesting that protection against the virus was maintained even after 30 years.

\begin{figure*}
\noindent\includegraphics[width=\ff\linewidth]{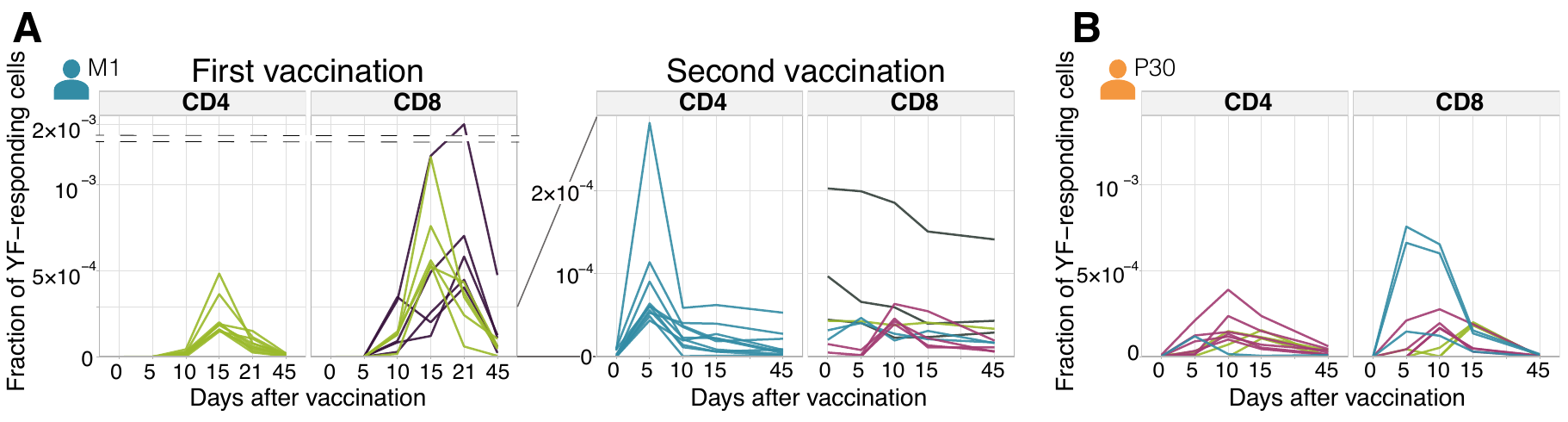}
\caption{{\bf Diversity of individual clonal trajectories in primary and secondary responses.  A, B.} Frequency of each YF-responding clonotype {\ch in bulk TCR repertoire} as a function of time. Individual clones show remarkable expansion after the primary response ({\bf A}, left panel) and secondary response both 18 months ({\bf A}, right panel) and 30 years ({\bf B}) after the primary vaccination. The ten most abundant (by peak frequency) CD4+ and CD8+ YF-responding clonotypes are shown for each vaccination. {\ch Clonal traces for all YF-responding clonotypes are shown in Fig. S3. }Color indicates the time of the response peak for each clonotype: blue for a peak at day 5, pink at day 10, green at day 15 and purple at day 21. Despite overall heterogeneity in clonal traces, more clones peak at early timepoints during the secondary response. {\ch Heterogeneity in clonal traces allows for expanded clones identification and computational alpha-beta TCR pairing (Fig. S6).}
}
\label{fig2}
\end{figure*}

\subsection*{\ch Diversity of clonal time traces in primary and secondary responses}

Our approach allows us to estimate the contribution of individual clones to the total response. 
We already showed  that the overall response strength to secondary immunization was an order of magnitude lower compared to the primary response. However, several clones showed remarkable expansion rates and peak frequencies, comparable to the ones observed in primary immunization. Such clones were observed in both donors upon secondary immunization after 18 months and 30 years (Fig. 2A and B, Fig. S3). We traced each single clone during primary and secondary response in donor M1. The concentration of clonotypes prior to the booster immunization correlated well (Pearson r=0.46 $p<0.0001$) with their concentration on day 45 after primary immunization (Fig. S4) suggesting a uniform contraction rate for all clones resulting in a half-life of 158$\pm$12.7 
days for the YF-specific T-cell subpopulation. Previously, Akondy et al. using deuterium labeling of cells specific to the immunodominant epitope NS4B$_{214-222}$
(as determined by a A02-NS4B$_{214-222}$-multimer binding assay) showed a very similar half-life of 123 days \citep{akondy_origin_2017}.

It was previously reported that only 5-6\% of YF-responding clones are preserved as immune memory, with the preferential recruitment of large clones \citep{dewitt_dynamics_2015}. By contrast, in our sample we could re-identify 96\% of CD4+ and 88\% of CD8+ clones that responded to the primary immunization in at least one sample after the booster immunization. This suggests that practically all the diversity of the responding repertoire is maintained in memory. The larger fraction of re-identified YF-responding clones in comparison to previous work may be explained by the sampling depth. Sequencing more T-cells will lead to the re-identification of even more YF-responding clonotypes.  

We then wanted to characterize how these persistent clonotypes responded to the booster vaccination. Interestingly, we found that the largest YF-specific CD8+ clones did not expand in response to the booster vaccine. Instead, the most expanded clonotypes were rare prior to the booster immunization (Fig. S5A). The situation was different for CD4+ cells: both high and low-frequency CD4+ clones expanded in response to the booster immunization (Fig. S5B).

{\ch The specific features of clonal trajectories shared by YF-responding clones make it possible to distinguish them from non-expanding clones, using unsupervised clustering (see Fig. S6AB and Methods). This method shows good concordance with edgeR and works also without biological replicates. 
In addition, we demonstrated that the heterogeneity of clonal trajectories could be leveraged to computationally pair alpha and beta chains from from bulk alpha and beta sequencing data, by exploiting the similarity of trajectories of alpha and beta clonotypes belonging to the same clone (see Fig. S6C and Methods).}

\begin{figure*}
\noindent\includegraphics[width=0.7\linewidth]{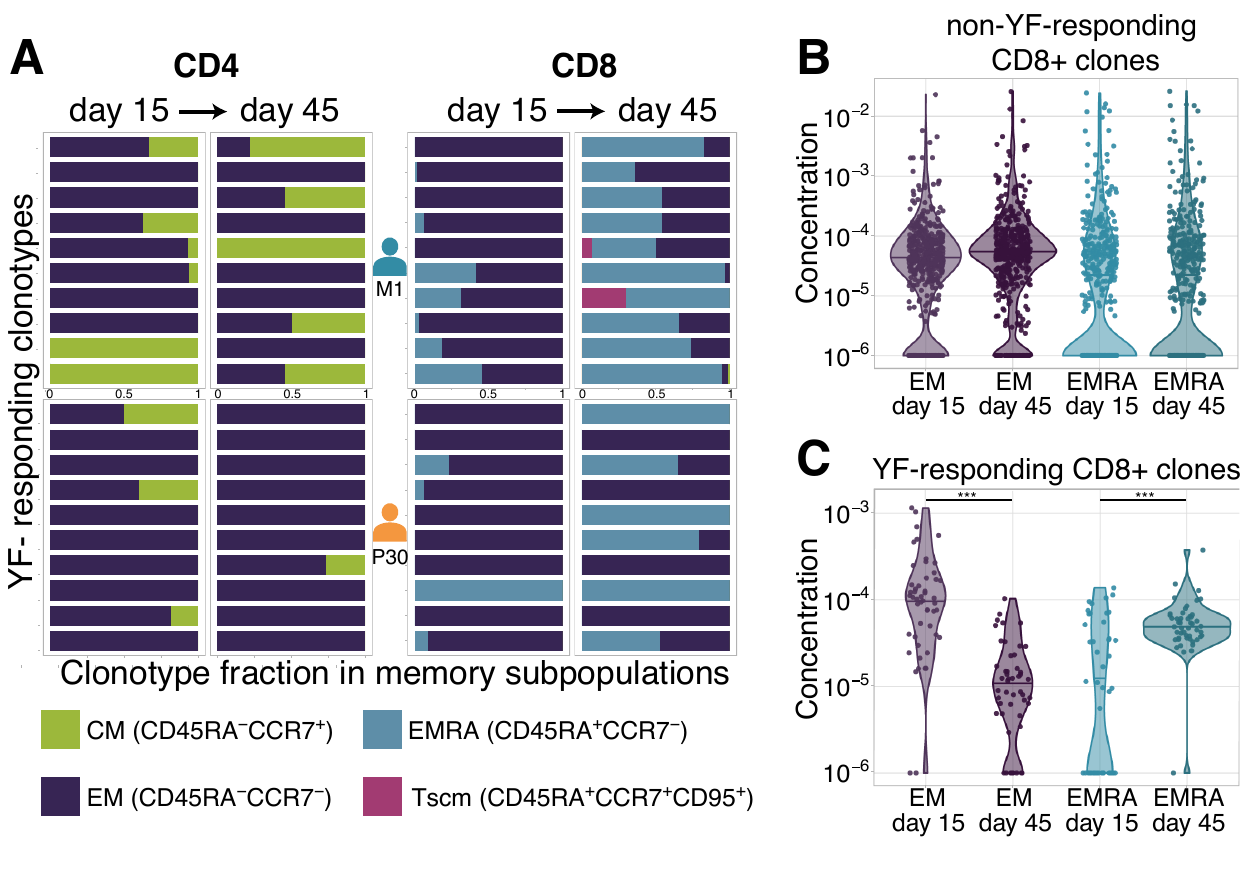}
\caption{{\bf Distribution of clonotypes in memory subsets. A.} Each color bar shows the estimated distribution of T-cell clones between memory subpopulations for a set of CD4+ (left panel) and CD8+ (right panel) clonotypes for donors M1 (top) and P30 (bottom) on day 15 and day 45. Each panel shows the 10 most abundant YF-responding clones in each donor on day 45, which are present in at least one memory subpopulation on both day 15 and day 45. {\bf B.} Estimated concentration of {\ch CD8+} clones with a given phenotype at different timepoints in the bulk PBMC repertoire, for non-YF-responding clonotypes and {\bf(C.)} YF-responding {\ch CD8+} clonotypes (Mann Whitney U-test, EM: p-value = $2.1\cdot 10^{-12}$, EMRA: p-value = $1.2\cdot 10^{-6}$). 
 Only clones with 30 or more Unique Molecular Identifiers (see Methods) in bulk repertoires on day 45 were used for the analysis. 
}
\label{fig3}
\end{figure*}

\subsection*{TCR sequencing shows the transition of clonotypes between memory subpopulations}
Several studies have reported subsets of long-lived memory YF-specific T-cells, whose concentration remained stable for years \citep{fuertes_marraco_long-lasting_2015,akondy_origin_2017}. It was shown that these long-lived memory cells are the progenies of effector cells, which divide vigorously during the peak of the response to the vaccine \citep{akondy_origin_2017}. TCR sequences can be used as ``barcodes'' to measure transitions between different memory subsets after YF immunization, defined by their surface markers revealed by flow cytometry.

We isolated with FACS (see Fig. S7 for the gating strategy) and sequenced TCR repertoires of 3 conventional T-cell memory subpopulations {\ch \citep{fuertes_marraco_long-lasting_2015,appay_phenotype_2008}}: effector memory (EM, CCR7-CD45RA-), effector memory re-expressing CD45RA (EMRA, CCR7-CD45RA+), and central memory  (CM, CCR7+CD45RA-) on days 0, 15, 45, and 18 months after the primary vaccination of donor M1 and on days 0, 15, and 45 after the second vaccination of donor P30. On day 45 we also isolated and sequenced the repertoire of the recently described Tscm (T-cell stem cell-like memory) subset (CCR7+CD45RA+CD95+).

On day 0, the concentration of almost all YF-responding clonotypes was too low to be detected in any of these subpopulations. However, we were able to calculate the distribution of YF-responding clonotypes between these phenotypes after immunization. {\ch In agreement with previous studies} the memory status of T-cell clones was tightly correlated with their CD4/CD8 status \citep{sathaliyawala_distribution_2013,thome_spatial_2014}. {\ch The vast majority of} CD4+ T-cell{\ch s}  were distributed between EM and CM, {\ch with $<1\%$ in EMRA}, while CD8+ T-cell clones were {\ch predominantly} found in EM and EMRA {\ch with $\sim 2\%$ in CM}. 
{\ch This difference also held for YF-responding clones (Fig. 3A).}
While for most CD8+ clonotypes in the total repertoire EM/EMRA phenotypes were stable between day 15 and day 45 (Fig. 3B, and Fig. S8A, C), the distribution of CD8+ YF-responding clones between memory subsets was significantly shifted towards the EMRA phenotype (Fig. 3C). This shift results from two processes: the rapid decay of EM cells (Fig. S8B) and the phenotype switch from EM to EMRA (Fig. S8D). Almost all YF-responding CD8+ clones detected 18 months after the first immunization corresponded to the EMRA phenotype (among 71 clones  found in more than 3 copies in bulk repertoire at day 0 before second vaccination, 41 were found only in the EMRA subset, 4 only in EM, and 6 in both). For CD4+ T-cells, we did not observe any trend in phenotype switching between days 15 and 45 after the vaccination. We hypothesize that switching from EM to CM phenotype was masked due to homing of CM cells to lymphoid organs, defined by the expression of the CCR7 chemokine receptor. 

\begin{figure*}
\noindent\includegraphics[width=\ff\linewidth]{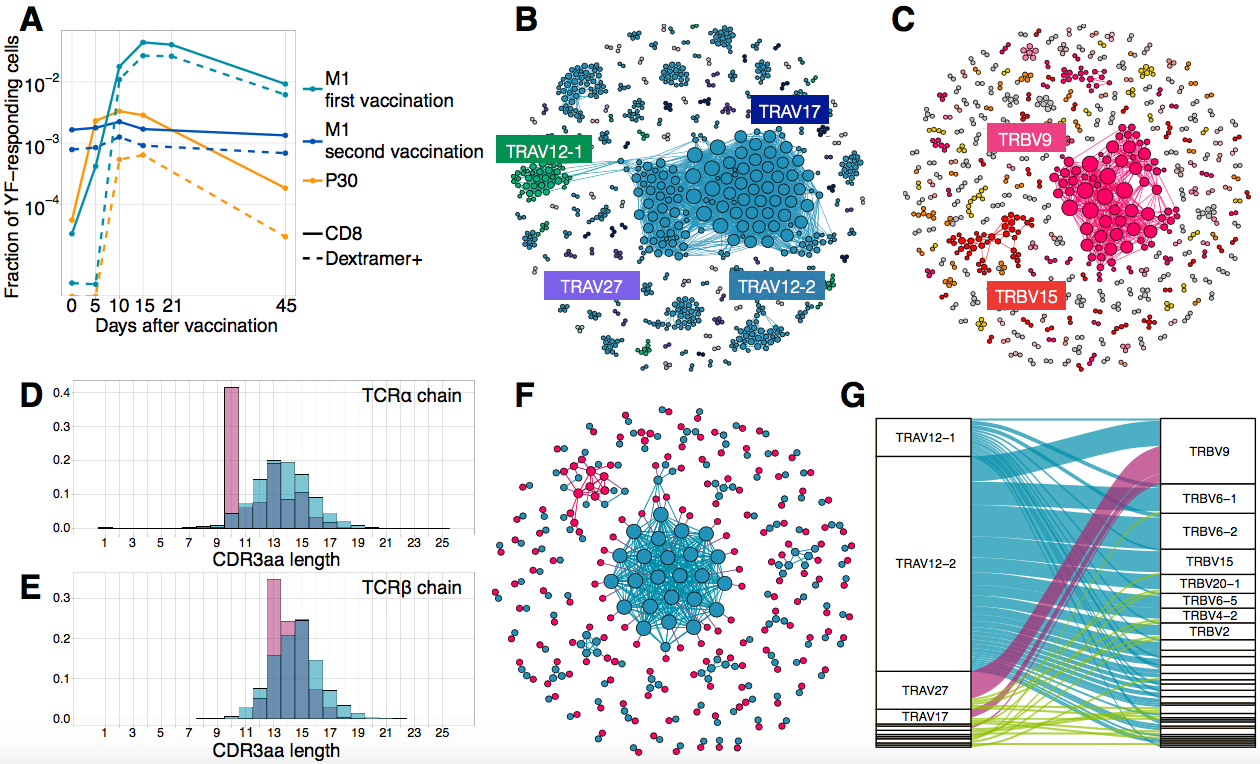}
\caption{{\bf Response to the immunodominant yellow fever epitope NS4B$_{214-222}$. A.} {\ch Fraction of all T-cells corresponding to} CD8+ YF-responding TCR$\beta$ clonotypes (solid lines) and CD8+NS4B-specific clonotypes (dashed lines) as a function of time post-vaccination (x-axis). Sequence similarity networks for TCR alpha {\bf(B)} and beta {\bf (C)} of NS4B-positive cells. Each vertex is a TCR amino acid sequence, connected with an edge if they differ by fewer than two mismatches. 
The size of the vertex indicates its degree. Vertices of zero degree are not shown. Color and text boxes indicate V-segments that are significantly enriched (exact Fisher test, Benjamini Hochberg adjusted $p<0.001$) in usage in epitope-specific cells compared to the bulk repertoire. NS4B-specific TCR alpha {\bf (D)} and TCR beta {\bf(E)} chains (red histograms) have biases in CDR3 length in comparison to bulk TCR repertoire of CD8+ cells (overlayed blue histograms). {\bf F.} Network of single-cell paired TCR alpha (blue) and TCR beta (red) of NS4B-specific TCRs. Vertices of the same color are connected if there are less than two mismatches in TCR chain amino acid sequence. An edge between vertices of different color represents the pairing of alpha and beta. The biggest alpha cluster (blue in the center) corresponds to the TRAV12-2 cluster on {\bf B}, and it pairs with many dissimilar beta chains. The biggest beta cluster (top left in red) corresponds to the TRBV9 cluster of {\bf C}. {\bf G.} Pairing of V-segments of TCR alpha (left) to V-segments of TCR beta (right) in scTCRseq of NS4B-specific T-cells. The height of each box is proportional to the number of unique clones with this V-segment. The width of ribbons is proportional to the frequency of TRAV-TRBV combination. NS4B-specific TCRs have two main binding modes, defined by TRAV12 segment family paired to almost any TRBV (blue) and by TRAV27 segment paired preferentially with TRBV9 (pink).
}
\label{fig4}
\end{figure*}

\subsection*{The response to a single immunodominant epitope can contribute to up to 60\% of the total response}
It was previously shown that in HLA-A02 donors the NS4B$_{214-222}$ LLWNGPMAV immunodominant epitope elicits the strongest CD8+ T-cell response \citep{akondy_yellow_2009,wieten_single_2016,kongsgaard_adaptive_2017,blom_temporal_2013}. Using an A02-pMHC-dextramer, we isolated NS4B-specific CD8+ T-cells from both donors (Fig. S9A,B) and applied TCR sequencing to get their unpaired TCR alpha and TCR beta repertoires. We obtained $\approx$2100 alpha  and $\approx$2000 beta functional receptor chains, one of the largest datasets for TCRs with a single specificity. YF-responding clonotypes identified by edgeR 
as expanded between timepoints are not restricted to any particular YF epitope and represent the repertoire targeted towards many different peptides presented by different HLA alleles. This allows us to quantify the relative contribution of NS4B-specific T-cells to the total anti-YF response. At the peak of the response, approximately 24\% of all YF-responding CD8+ T-cells were specific to NS4B in the donor vaccinated 30 years ago (P30), and up to 60\% in the first time vaccinee (M1) (Fig. 4A). However, NS4B-specific clonotypes could not be distinguished from other YF-responding clonotypes from their time traces alone, as they both responded with similar dynamics (Fig. S10).

\subsection*{Sequence analysis and structural modeling of NS4B-specific TCRs reveals two motifs with distinct peptide binding modes}
We next asked whether there are distinct features in the sequence of NS4B-specific TCRs, which might explain the immunodominance of this epitope. Figures 4B and C show sequence similarity networks for TCR alpha and TCR beta chains of NS4B-specific clonotypes. The TCR alpha repertoire shows biased V-usage and complementarity determining region 3 (CDR3) lengths (Fig. 4D).  TRAV12-2, TRAV12-1, TRAV27, and TRAV17 gene usage were significantly enriched in the NS4B-specific TCRs (exact Fisher test, Benjamini Hochberg adjusted $p<0.001$), with more than 45 percent of the clonotypes expressing TRAV12-2, in comparison to just 4.5\% of TRAV12-2 in the total CD8+ TCR repertoire. Beta chains formed several distinct clusters of highly similar sequences, with significant but less marked V-usage biases towards TRBV9, TRBV15, and TRBV6-1/2, as well as some bias in the length distribution (Fig. 4E). Almost 37\% of NS4B-specific clonotypes used TRBJ2-7.

We next asked how these clusters of highly similar sequences in the alpha and beta NS4B-specific repertoires corresponded to each other. Prior to booster immunization, we isolated NS4B-specific T-cells from donor M1 (Fig. S9C) and performed single-cell RNA sequencing (scRNAseq) and single-cell paired TCR sequencing (scTCRseq). We collected data from 3500 cells corresponding to 164 clonotypes (see Methods). Fig. 4F shows a joint similarity network for TCR alpha and TCR beta chains, with both intra-chain sequence similarity and inter-chain pairings. Alpha-beta pairing seemed to be mostly random, with some exceptions: for instance, specific TCRs using the most dominant TRAV12-2 alpha motif were paired with many different beta chains with a broad usage of V-segments (Fig. 4G and Fig. S11A), but with a restricted CDR3$\beta$ length of 13--14 amino acids. TCRs using TRAV27 and TRBV9 segments were also preferentially paired with one another (Fig. S11C). Clustering of paired sequences using the TCRdist measure (Fig. S11B) resulted in two large clusters corresponding to these two major motifs with conserved V-usage.

The preferential usage of the TRAV12 family was reported before for TCRs responsive to the NS4B epitope \citep{bovay_t_2018,zhang_high-throughput_2018}. It was speculated \citep{bovay_t_2018}, that the CDR1$\alpha$ of this V-segment forms contacts with the peptide. To test this hypothesis, we modeled the 3D structures of clonotypes from scTCRseq using the Repertoire Builder server \citep{schritt_repertoire_2019} and then docked the resulting model structures using RosettaDock \citep{lyskov_rosettadock_2008} to the HLA-A02 pMHC complex structure, recently solved using X-ray crystallography \citep{bovay_t_2018}, see Methods for details. 
Models of TCR-pMHC complexes 
showed  that the TRAV12-2 TCRs formed more contacts with the peptide using  CDR1$\alpha$ loops, and fewer contacts with CDR3$\alpha$ loops, in comparison to TRAV27 TCRs (Fig. S12A). Interestingly, CDR3$\alpha$ sequences of TRAV12-2 TCRs were very similar to the ones observed in the repertoire of the same donor prior to the immunization, suggesting absence of epitope-driven selection of the CDR3$\alpha$ of these TCRs (Fig. S12B). Based on these results, we hypothesize that TCRs using TRAV12 and TRAV27 motifs represent two independent and distinct solutions to the binding of the NS4B epitope.

\begin{figure*}
\noindent\includegraphics[width=0.9\linewidth]{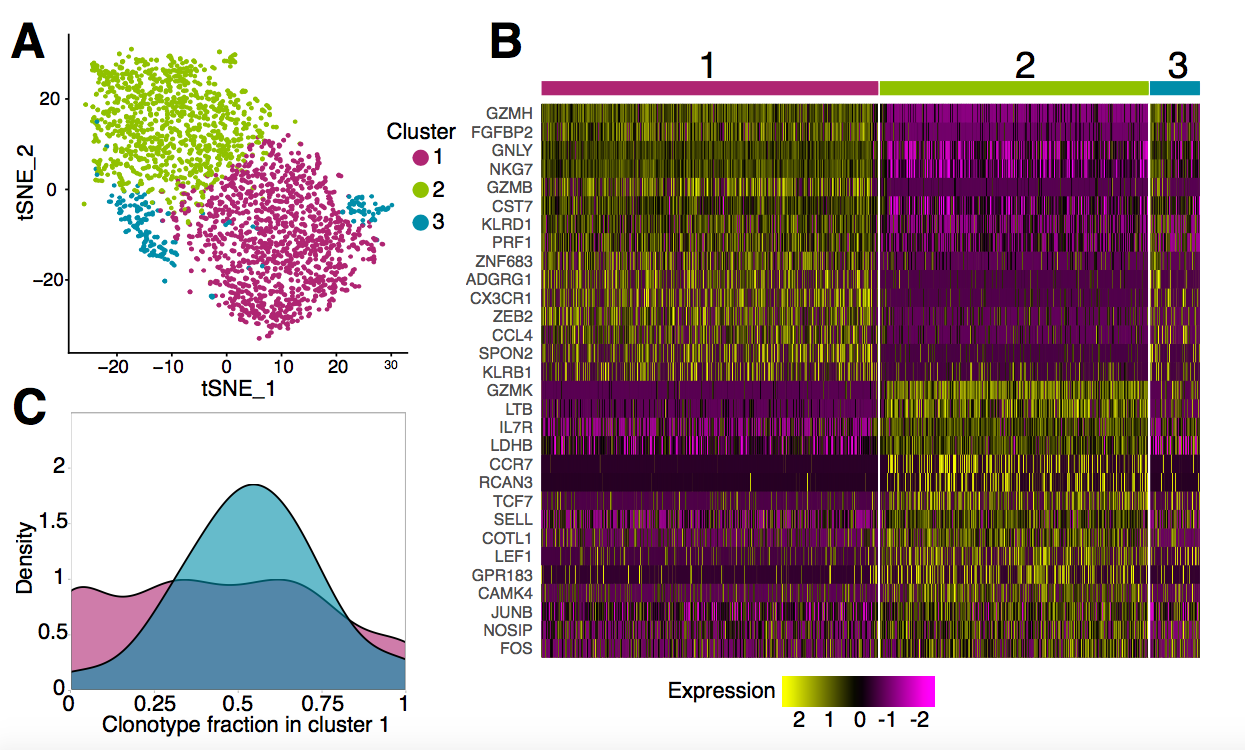}
\caption{{\bf Phenotypic diversity of NS4B-specific cells 18 months after yellow fever immunization. A. } 2D t-SNE visualization of unsupervised clustering (Seurat analysis) of RNAseq data based on 2000 most variable genes shows three distinct clusters of NS4B-specific cells. {\bf B.} The heatmap of top 15 significantly enriched genes of single cells in clusters 1 and 2 defined by the MAST algorithm. The panel above the heatmap identifies the cluster identity of the cells. {\bf C.} Gaussian kernel density estimate for the relative fraction of cells belonging to cluster 1 for each clonotype. Blue distribution shows the theoretical prediction under the null hypothesis: clonotype labels were shuffled between cells (1000 permutations). The observed distribution is flatter than the theoretical one, indicating the presence of clonotypes with either a minority or a majority of cells belonging to cluster 1 (${\chi}^2$-test with MC-estimated p-value=0.0005).
}
\label{fig5}
\end{figure*}

\subsection*{scRNAseq of NS4B-specific T-cells reveals two distinct cytotoxic phenotypes}
Next we used the scRNAseq gene expression data to investigate the phenotype of specific T-cells in finer detail.
While almost all NS4B-specific clonotypes 18 months after  vaccination belonged to the conventional EMRA subset, scRNAseq revealed huge heterogeneity of gene expression inside this population.
Unsupervised clustering by Seurat 3.0 software  \citep{stuart_comprehensive_2019,butler_integrating_2018} (see Methods) revealed three sub-phenotypes of NS4B-specific cells (Fig. 5A). 

Overall we found 166 genes that were differentially expressed according to the MAST algorithm \citep{finak_mast:_2015} between these clusters (Fig. 5B). Cells from cluster 1 showed high expression of cytotoxicity related genes \textit{GZMB, GNLY, GZMH, NKG7, PRF1, CX3CR1, SPON2, KLRD1}, Hobit and T-bet transcription factors (Fig. S13A). The combination of these genes also suggests that this cytotoxicity is mediated by the perforin pathway. The second cluster of cells is enriched in  genes such as \textit{CCR7, TCF7, SELL, JUNB, LEF1,} and especially \textit{IL7R} which are essential for long-term survival and maintenance of memory T-cells (Fig. S13B) \citep{jeannet_essential_2010,zhou_differentiation_2010,kaech_selective_2003,jung_ccr7_2016,schluns_interleukin-7_2000}. However, these cells also express unique markers related to cytotoxicity: \textit{GZMK, LTB} as well as \textit{KLRG1, KLRB1}, T-bet, and \textit{GZMH,} albeit at lower levels than cells in cluster 1.

Very similar clusters of genes were found in single-cell RNAseq analysis of CD4-cytotoxic lymphocytes EMRA cells \citep{patil_precursors_2018}. The expression pattern of granzymes and killer-like receptors in our clusters suggests that cells in cluster 2 may be the precursors of cells in cluster 1. The expression of \textit{GZMK} (enriched in cluster 2) was shown to be prevalent in early memory stages \citep{harari_distinct_2009,bratke_differential_2005}, while high levels of \textit{GZMB, GZMH, KLRB1, KLRG1,} and \textit{ADGRG1} (enriched in cluster 1) are associated with more terminally differentiated memory cells with higher cytotoxic potential \citep{truong_killer-like_2019,takata_three_2006}. Interestingly, cluster 2 has {\ch higher} expression of genes encoding ribosomal proteins, which were recently reported to be a feature of memory precursor cells \citep{araki_translation_2017}. The transition of cells between the two clusters is also supported by the existence of cluster 3, which shows intermediate gene expression of cluster 1 and 2 markers, and thus may represent cells gradually changing phenotype.

For each cell from the scRNAseq experiment, we obtained matched scTCRseq results. We wondered whether the TCR clonotype influenced cell gene expression profile. Interestingly, the distribution of clonotypes between clusters was not random (${\chi}^2$-test with MC-estimated p-value=0.0005): some clonotypes showed a clear preference for one of the phenotypes (Fig. 5C). To match single-cell gene expression data with measurements of clonotype concentrations obtained with TCRseq, we averaged mRNA counts over the all cells of the same clonotype, and repeated the differential gene expression analysis (see Methods). 
We obtained two clusters of clonotypes with the same enriched genes (Fig. S14A) as observed for clusters of single cells (Fig. 5B), confirming the association of phenotype and clonotype. Clonotypes from both clusters expanded following the second immunization, indicating that both phenotypes are capable of response. Clonotypes associated to cluster 1 had larger frequencies both on day 45 after the first vaccination (Fig. S14B, left), and 18 months later before the booster shot (Fig. S14B, right), than clonotypes associated to cluster 2. This result suggests that even for T-cells recognizing the same epitope, particular clones are linked to particular memory phenotype.

\section*{Discussion}

In this study, we applied high-throughput sequencing of TCR alpha and TCR beta repertoires to track T-cell immune response to primary and secondary immunization with yellow fever vaccine. This approach does not require previous knowledge of TCR specificity and thus allows us to quantify and compare the response of individual T-cell clones recognizing different epitopes on the same scale. 

We found that up to 60\% of all responding CD8+ T-cells were specific to a single immunodominant peptide. Several studies reported high precursor frequency of T-cells reactive to this epitope \citep{zhang_high-throughput_2018,bovay_t_2018}. Bovay et al. recently suggested that recognition of antigenic peptide through the germline-encoded CDR1 loop of the TRAV12 segment is one of the main reasons for high precursor frequency \citep{bovay_t_2018}. This hypothesis is supported by our TCR structural modeling and TCR-pMHC docking simulations, as well as by the analysis of the NS4B-specific T-cell repertoire. We also identified an additional motif defined by TRAV27+TRBV9+ TCRs. 
It will be interesting to investigate if these two motifs differ in binding affinity or are susceptible to potential escape mutations that can appear in the antigenic peptide. Another question is how diverse is the level of clonal response to the YF vaccine  in HLA-A02 negative donors, and what fraction of the response is directed towards the most immunodominant epitopes in the context of other HLA types.

Most previous studies focused on TCR beta repertoires, partially because the diversity of TCR beta is higher, making it a better marker for clonal tracking \citep{chu_longitudinal_2019}. We found that TCR alpha may be used for clonal tracking as well, giving almost the same results as TCR beta in terms of the number of expanding clonotypes and their cumulative fractions on different timepoints. In the particular case of the response of HLA-A02 donors to the YF vaccine, the TCR alpha repertoire turned out to be even more informative, as T-cells responding to the immunodominant epitope predominantly use certain TRAV segments.

One of the major limitations of bulk TCR sequencing is that the resulting repertoires are unpaired, while TCR specificity in most cases is defined by the combination of alpha and beta chains. We show that the simultaneous sequencing of bulk alpha and beta repertoires performed on many timepoints allows us to make predictions on alpha-beta pairing. Even with the rise of single-cell sequencing, this method might still be of interest since most available single-cell platforms can only analyze limited numbers of $10^4$-$10^5$ cells. In addition, these experiments are still expensive in comparison to the bulk TCR sequencing, which enables the profiling of millions of lymphocytes more cheaply.  

We found that $\approx 90$\% clonotypes responding to primary immunization were present in peripheral blood 18 months after immunization. Recently, Akondy et al. showed using deuterium cell labeling that long-survived memory cells have a history of intense clonal expansion, and thus are likely to differentiate from effector cells after response \citep{akondy_origin_2017}. This explains both the high remaining diversity of YF-responding clonotypes and a proportional decrease of these clonotypes between primary and booster immunizations.

Interestingly, we observed a very different response of CD4+ and CD8+ memory cells to the booster vaccination. It may be explained by differences in antigen presentation mechanisms: CD4+ T-cells may be activated well by antigen presenting cells phagocyting neutralized viral particles and presenting exogenous peptides on MHC-II complexes, while CD8+ memory cells can be more efficiently triggered by a productive viral infection resulting in the presentation of endogenously translated viral proteins on MHC-I. It was previously shown that the magnitude CD8 response depends on the viral load \citep{akondy_initial_2015}.

It will be interesting to perform a similar study in donors vaccinated with YF backbone chimeric vaccines, where genes from other viruses substitute some of the YFV17D genes. It was shown that preexisting anti-YF immunity \citep{monath_clinical_2002} does not affect the formation of neutralizing antibodies to the novel virus. This finding suggests that not only efficient reactivation of existing CD4 memory but also the formation of CD4 responses to novel epitopes is possible during the booster with slightly different antigen. 

We found that, while the overall secondary response to the vaccine was much smaller both in terms of clonal diversity and cumulative frequency, a few clones still undergo strong clonal expansion. This may be indirect evidence for the programmed proliferation hypothesis \citep{moore_dependence_2019} according to which a single encounter of a TCR with an antigen triggers several rounds of T-cell division. It was shown that the virus is undetectable in the peripheral blood after booster vaccination \citep{reinhardt_development_1998}, meaning that the amount of available antigen is much lower, and so is the number of encounters and responding clonotypes.  

The transition between EM and EMRA phenotypes in CD8+ clones responding to yellow fever vaccine was previously measured using flow cytometry \citep{wieten_single_2016,fuertes_marraco_long-lasting_2015}. Here we confirm these reports with high-throughput sequencing, using TCR as a barcode to mark cells of the same clonal lineage. Furthermore, we identified two distinct cytotoxic phenotypes in NS4B-specific T-cells 18 months after primary immunization. It is not clear why the distribution of clonotypes between two these phenotypes was biased. Since we performed scRNAseq of clonotypes specific to the single antigen, these differences might be either the consequence of different TCR affinity or some phenotypic heterogeneity present in the precursor cells. Additional experiments at later timepoints would be required to estimate the longevity of these clonotypes. 

To summarize, we show that vaccination with YFV17D leads to the recruitment of a diverse repertoire of T-cells, which is then available as immune memory for the secondary response years after the immunization. Even T-cells with the same epitope specificity show several distinct motifs in TCR and have different memory phenotypes. Such heterogeneity of cells might be crucial for individual immune response robustness, underlying cross-reactive responses to similar viruses, and the possibility to escape mutants, which could be tested directly in future studies. However, this diverse T-cell response is strongly focused on single HLA-A02 restricted epitope. An interesting question is how many distinct foci of response exist in the human population with a variety of HLA-types; and how this diversity of individual responses contribute to the defense from the infection at the population level. Systematic studies of donors with different genetic backgrounds and corresponding immunodominant epitope-specific repertoires will be able to address this question.

\section*{Methods}
\subsection*{Donors and blood samples}
Blood samples were collected from two healthy donors (M1 male age 26, and P30 male age 39) on multiple timepoints before and after immunization with YFV17D vaccine. All donors gave written informed consent to participate in the study under the declaration of Helsinki. The blood was collected with informed consent in a certified diagnostics laboratory. The study was approved by the institutional review board (IRB) of Pirogov Russian National Research Medical University. HLA haplotypes of donors (Table S1) were determined by in-house RNA-based amplification and sequencing method.
\subsection*{Isolation of T-cell subpopulations}
We isolated PBMCs from the blood using standard Ficoll-Paque protocol. CD4 and CD8 fractions were isolated with CD4/CD8 Positive Selection Dynabeads Kits according to the manufacturer's protocol. For isolation of memory subsets, we stained PBMCs with the mix of antibodies: anti-CD3-FITC (UCHT1, eBioscience), anti-CD45RA-eFluor450 (HI100, eBioscience), anti-CCR7-AlexaFluor647 (3D12, BD Pharmingen), anti-CD95-PE (DX2, eBioscience). Four subsets of cells were sorted into RLT buffer (Qiagen) on BD FACS Aria III: EM (CD3+CD45RA-CCR7-), EMRA (CD3+CD45RA+CCR7-), CM (CD3+CD45RA-CCR7+), Tscm (CD3+CD45RA+CCR7+CD95+). HLA-A02 dextramer loaded with the NS4B$_{214-222}$ peptide (LLWNGPMAV) from YFV17D (Immudex) was used for epitope-specific T-cells isolation. Cells were stained with NS4B-dextramer-PE, anti-CD3-eFluor450 (UCHT1, eBioscience), and anti-CD8-FITC (SK1, eBioscience) according to the manufacturer's protocol. RNA was isolated using standard TriZol protocol (for bulk PBMCs, CD4 and CD8, NS4B-specific and negative fractions) or RNAeasy Micro Kit (Qiagen) (for memory subsets). The amount of RNA was measured on Qubit 2.0 (Invitrogen). 
\subsection*{Sample preparation for the single-cell gene expression and immune profiling}
For 10x Genomics single-cell gene expression and immune profiling, we used PBMCs isolated from 60 ml of blood of donor M1 before the second immunization. PBMCs were stained with NS4B-dextramer-PE (Immudex) according to the manufacturer's protocol. Additionally, cells were stained with anti-CD3-eFluor450 (eBioscience), and anti-CD8-FITC (eBioscience). Previous to FACS sorting procedure, we used propidium-iodide to mark dead cells. As the NS4B-specific cell frequency was very low (Fig. S9C), we used anti-PE Ultra-pure MicroBeads (Miltenyi) for the enrichment. In brief, every milliard of PBMCs was incubated with 10 $\mu$l of magnetic beads for 15 minutes on ice. After a washing step with PBS 5\% FCS, the cell suspension was applied on MS MACS Column (Miltenyi). Columns were washed three times with PBS 5\% FCS and stained with propidium-iodide just before the FACS (FACS Aria II). This procedure resulted in a dramatic increase of NS4B-specific cell frequency in the sample (Fig. S9C) and accordingly lead to reduced FACS procedure time. 
For single-cell immune profiling of bulk T-cell clonotypes from PBMCs, we stained the cells with anti-CD3-eFluor450 (Invitrogen) and propidium-iodide, thus selecting CD3 positive cells. Approximately 10,000 CD3+ cells were used for 10x Genomics VDJ T-cell receptor enrichment protocol.
\subsection*{High throughput T-cell repertoire sequencing}
Libraries of TCR alpha and TCR beta chains were prepared as previously described \citep{pogorelyy_persisting_2017}. In brief, isolated RNA was used for cDNA synthesis with 5'RACE template switch technology to introduce universal primer binding site and Unique Molecular Identifiers (UMI) at the 5' end of RNA molecules. Primers complementary to both TCR alpha and TCR beta constant segments were used for cDNA synthesis initiation. cDNA was amplified in two subsequent PCR steps. During the second PCR step, sample barcodes and sequence adapters were introduced to the libraries. Libraries for the fractions with low amount of cells were prepared using SMART-Seq v4 Ultra Low Input RNA kit (TakaraBio). Libraries were sequenced on Illumina platform HiSeq 2500 with 2$\times$100 bp sequencing length or NovaSeq 2$\times$150 bp sequencing length. 
Parallel single-cell alpha/beta TCR and 5' gene expression sequencing was performed using 10x Genomics Kits (Chromium Single Cell A Chip Kit, Chromium Next GEM Single Cell 5' Library and Gel Bead Kit, Chromium Single Cell V(D)J Enrichment Kit, Human T Cell, Chromium Single Cell 5' Library Construction Kit, Chromium i7 Multiplex Kit) according to the manufacturer's protocol. Libraries were sequenced on Illumina platform HiSeq 3000 with 2$\times$150 bp sequencing length. 
\subsection*{Repertoire data analysis}

{\bf Raw data preprocessing.} Raw repertoire sequencing data were preprocessed as described in \citep{pogorelyy_persisting_2017}. Briefly, sequencing reads were demultiplexed and clustered by UMI with MIGEC software \citep{shugay_towards_2014}. The alignment of genomic templates to the resulting consensus sequences was performed with MiXCR \citep{bolotin_mixcr:_2015}. Raw sequencing data obtained from RNAseq experiments were analyzed directly with MiXCR using default RNAseq analysis pipeline. 

{\bf Identification of changed clonotypes by edgeR.} To identify TCR alpha and TCR beta clonotypes that significantly expand after YF vaccination, we used the edgeR package \citep{robinson_edger:_2010} as previously described \citep{pogorelyy_precise_2018}. In brief, for each timepoint, we used two biological replicates of bulk PBMC. TMM-normalization and trended dispersion estimates were performed according to edgeR manual. We used an exact test based on the quantile-adjusted conditional likelihood (qCML) to identify clonotypes significantly expanded between pairs of timepoints. A clonotype with {\ch FDR adjusted p-value$<0.01$ (exact qCML-based test)} was considered YF-responding if its log$_2$-fold change estimate log$_2$FC$>5$ between any pairs of timepoints from 0 to the peak of the primary response (day 15). {\ch The usage of log$_2$FC$>5$ threshold in addition to p-value threshold is important to filter statistically significant but small clonal expansions, which were previously shown to occur in healthy donors in the absense of vaccination on the timescale of one week, see \citep{pogorelyy_precise_2018}. } 
CD4/CD8 \textit{in silico} phenotyping was performed as suggested before \citep{pogorelyy_precise_2018}: for each clone from bulk PBMC repertoire we assign CD4 phenotype if it is more abundant in the sequenced CD4 repertoire and \textit{vice versa}.  {\ch Over $98\%$ of clonotypes were found exclusively in CD4 or CD8 compartment. However, a small group of clonotypes ($1.4\%$ for TCR alpha and $0.14\%$ for TCR beta for day 15 timepoint of donor M1) was present in both compartments in comparable frequencies. These clonotypes have significantly higher TCR generative probabilities than others ($p<0.001$, Mann Whitney U-test) and thus are likely to arise from convergent recombination of the same TCR chain in both compartments. }

{\ch To quantify the magnitude of the response on each timepoint we inferred the fraction of YF-responding cells as the proportion of all $\alpha\beta$T-cells. To estimate this quantity from TCR repertoire data, for each susbset of interest (CD4+, CD8+, or NS4B-specific YF-responding clonotypes) we calculate the cumulative frequency of these clonotypes in TCR repertoire of bulk PBMCs in each timepoint.}

{\bf Identification of YF-responding clonotypes by Principal Component Analysis (PCA).} We chose clonotypes that appeared in the top 1000 most abundant clonotypes at any timepoint after primary immunization. For these clonotypes, we made matrices of frequencies on all timepoints after primary immunization. Before applying PCA to these matrices, each value was normalized by dividing on maximum frequency for this clonotype. For cluster identification, we used hierarchical clustering with average linkage on euclidean distances between clonotypes. The number of clusters was set to 2. This analysis was performed for both alpha and beta chains of donor M1. For the twin donors \citep{pogorelyy_precise_2018}, only replicate F1 was used for {\ch expanded} clones identification.

{\bf Memory transition analysis.} For this analysis, we used clonotypes that had at least 30 UMIs at day 45 after primary vaccination.  The clonotype frequency in memory subset is multiplied by the number of cells obtained by FACS on this timepoint for this subset. Then adjusted frequencies are normalized across all subsets to get a partition of each TCR clonotypes across subsets. Obtained partitions were multiplied by the frequency of a clonotype in bulk at this timepoint to get the concentration of clonotypes with a particular memory phenotype in the bulk repertoire. 

{\bf Computational decontamination of NS4B-specific repertoire.} Since FACS sorting is not precise, TCR repertoires of the population of interest often contains abundant clonotypes from the bulk population. To obtain a list of NS4B-specific TCRs we took clonotypes that were enriched (at least 10 times) in the A02-NS4B-dextramer positive fraction compared to A02-NS4B-dextramer negative fraction. We also discarded TCR clonotypes that were more abundant in CD4 than CD8 subpopulation on day 0 (as only CD8 cells should bind to A02 which is a MHC I allele). {\ch Although $\sim 30 \%$ of resulting unique NS4B-specific clonotypes overlapped with the list of significantly expanded clonotypes identified with edgeR, they corresponded to $\sim 90\%$ of NS4B-specific T-cells.} 

\subsection*{Computational pairing of TCR alpha and TCR beta from bulk repertoires}
For pair of clonal time traces we used a Euclidian distance between transformed frequencies:

$$
D(C_{\alpha},C_{\beta})=\sqrt{{\sum_{i}{(t(C_{\alpha,i})-t(C_{\beta,i})})}^2},
$$
where $C_{\alpha,i}$ and $C_{\beta,i}$ are the concentrations of an $\alpha$ and a $\beta$ chain on the $i$-th timepoint. The transformation $t$ of clone concentration $C$ was chosen to address the overdispersion of frequencies at large concentrations (see \citep{pogorelyy_precise_2018}):
$$
t(C_i)=\log_{10}{(\sqrt{a+bC_i}+\sqrt{bC_i})},
$$
where $a=4.26\times10^{-6}$ and $b=3.09\times10^{-3}$.
To address possible systematic bias in expression between $\alpha$ and $\beta$ chains in a clonotype, we introduce a log-fold shift $\lambda$ in a trajectory with a quadratic penalty ($\mu$=0.1):
$$
D_s(C_{\alpha},C_{\beta})=\min_{\lambda}{(D(C_{\alpha},10^{\lambda}C_{\beta})+\mu\lambda^2)}.
$$
We calculated $D_s$ distances between each pair of $\alpha$ and $\beta$ clonotypes out of the 1000 most abundant ones in the bulk repertoires on day 15 post-vaccination. For each $\alpha$ clonotype, we picked the 5 closest $\beta$ clonotypes as candidate pairings.
{\ch
As a benchmark, we used two single-cell TCR sequencing (scTCRseq) experiments using the 10x Genomics platform and obtained paired repertoires for samples of bulk T-cells (CD3+) and YF epitope-specific T-cells (CD8+NS4B-dextramer+). Note that these two samples are very different in their clonal time traces: NS4B-specific clones show very active response dynamics, expanding and contracting in the course of primary and booster immunization, while the CD3+ T-cell sample corresponds to the most abundant clones in the repertoire, which are largely stable between timepoints. 
A $\alpha\beta$TCR clonotype from 10x genomics experiment was considered correctly paired from bulk TCRseq data using the algorithm if the correct TCR beta was present among the 5 most probable TCR beta sequences predicted for the TCR alpha of this clonotype. Out of the 62 NS4B-specific clonotypes sampled in the 10x Genomics experiment, we were able to computationally identify 41 correct pairs from the bulk TCRseq data. Out of 26 CD3+ T-cell clonotypes, 20 were paired correctly.  }

\subsection*{Paired single-cell TCR sequencing}
To investigate TCR chains pairing in YF-specific clonotypes, we performed single-cell immune profiling with 10x Genomics protocol. The analysis of the data with Cell Ranger 2.2.0 (10x Genomics) with default parameters resulted in 3244 cells corresponding to 986 clones. Many of these clones had multiple TRA/TRB chains and are likely to represent multimers of cells (Fig. S15A). For further analysis, we chose only high-confident clones that had one TRA and one TRB chain and were present more than twice in the data. This procedure resulted in the list of $\approx 2000$ cells corresponding to 164 TCR alpha/beta clones.

\subsection*{TCR-pMHC complex modeling}
Models for each paired alpha-beta TCRs from 10x Genomics data were constructed using the RepBuilder server  (\url{https://sysimm.org/rep_builder/}) \citep{schritt_repertoire_2019} , and then docked to HLA-A02-LLWNGPMAV complex using rosettaDock2 (\url{https://www.rosettacommons.org/software}) routine \citep{lyskov_rosettadock_2008}. 152 TCRs passed the modeling step. For each TCR we obtained 1000 decoys in docking simulations. The thirty best decoys (by interface score) were used to calculate a contact map with the bio3d R package \citep{grant_bio3d:_2006}. It was previously shown \citep{pierce_flexible_2013}, that some docking decoys exhibit binding modes which are not found in natural TCRs. In the analysis, we only used decoys in which the root mean squared deviation between the centers of mass of the alpha and beta chains in the decoys, and the centers of mass of these chains in at least one published HLA-A02-TCR complex from ATLAS database \citep{borrman_atlas:_2017}, were less than 4 \si{\angstrom}. 
The number of contacts to the peptide was averaged over decoys that passed the threshold. Only clonotypes with $\geqslant 5$ of resulting filtered decoys were used for the analysis (see Fig. S12A). 

\subsection*{Single cell gene expression analysis}
For single-cell gene expression analysis, we pre-processed the data with Cell Ranger 2.2.0 (10x Genomics). We used GRCh38-1.2.0 reference genome for the gene alignment. The resulting gene count matrix was analyzed with Seurat 3.0 package \citep{stuart_comprehensive_2019,butler_integrating_2018}. Cells that had fewer than 200 features detected were filtered out. We also filtered out features that were present in fewer than 3 cells 
and genes of TCR receptors (e.g., \textit{TRAV, TRAJ, TRBV, TRBJ}), as they are the source of unwanted variation in the data (Fig. S15B). Then a standard data pre-processing was performed to remove low-quality cells and cells multiplets. We filtered out cells that had more than 2700 features or more than 8\% of mitochondrial genes (Fig. S15C). Feature expression measurements for each cell were normalized using default log-normalization in the Seurat package. Following the manual's suggestion, the 2000 most variable features were selected for further analysis. Prior to dimensionality reduction, data were scaled so that the mean expression was {\ch 0} and the variance equals to 1. 
The first 10 dimensions of PCA were used for cluster identification with the resolution parameter set to 0.4. To identify differentially expressed genes between clusters we used the MAST algorithm \citep{finak_mast:_2015} implemented in the Seurat package. We only tested genes that were present in more than 25\% of cells in any group and that had at least a 0.25 log fold difference between the two groups of cells. 

We performed a similar analysis to identify differentially expressed genes between clonotypes (rather than individual cells). We created a matrix containing the mean gene expressions over cells within each clonotype, and treated it like normal single-cell results. In this case, we did not filter multiplet cells (with a high number of features and a high percentage of mitochondrial genes), as all our ``cells'' were indeed computational multimers. The rest of the analysis was performed in the same way. To check the results we shuffled cell barcodes between the clonotypes and repeated the analysis. All cells ended up in a single cluster for this random control. 

\section*{Acknowledgments}
{\ch We thank J.C. Crawford and P.G. Thomas for assistance with TCRdist software and for helpful discussions.}
This work was funded by the European Research Council Consolidator Grant n. 724208 and RSF 15-15-00178. 
IZM was supported by RFBR 18-29-09132 and 19-54-12011. PB, ER and AF were supported by the Deutsche Forschungsgemeinschaft (DFG) through the Cluster of excellence Precision Medicine in Chronic Inflammation (Exc2167). ER was partially supported by DFG 4096610003. DMC was supported by grant 075-15-2019-1660 from the Ministry of Science and Higher Education of the Russian Federation to the Center for Precision Genome Editing and Genetic Technologies for Biomedicine under Federal Research Programme for Genetic Technologies Development from 2019 to 2027.

\bibliographystyle{pnas}

\begin{thebibliography}{10}

\bibitem{murugan_statistical_2012}
Murugan A, Mora T, Walczak AM, Callan CG
\newblock (2012) Statistical inference of the generation probability of
  {T}-cell receptors from sequence repertoires.
\newblock \emph{Proceedings of the National Academy of Sciences of the United
  States of America} 109:16161--16166.

\bibitem{miller_human_2008}
Miller JD, {et~al.}
\newblock (2008) Human effector and memory {CD}8+ {T} cell responses to
  smallpox and yellow fever vaccines.
\newblock \emph{Immunity} 28:710--722.

\bibitem{akondy_yellow_2009}
Akondy RS, {et~al.}
\newblock (2009) The yellow fever virus vaccine induces a broad and
  polyfunctional human memory {CD}8+ {T} cell response.
\newblock \emph{Journal of Immunology (Baltimore, Md.: 1950)} 183:7919--7930.

\bibitem{akondy_initial_2015}
Akondy RS, {et~al.}
\newblock (2015) Initial viral load determines the magnitude of the human {CD}8
  {T} cell response to yellow fever vaccination.
\newblock \emph{Proceedings of the National Academy of Sciences of the United
  States of America} 112:3050--3055.

\bibitem{blom_temporal_2013}
Blom K, {et~al.}
\newblock (2013) Temporal dynamics of the primary human {T} cell response to
  yellow fever virus 17d as it matures from an effector- to a memory-type
  response.
\newblock \emph{Journal of Immunology (Baltimore, Md.: 1950)} 190:2150--2158.

\bibitem{kohler_early_2012}
Kohler S, {et~al.}
\newblock (2012) The early cellular signatures of protective immunity induced
  by live viral vaccination.
\newblock \emph{European Journal of Immunology} 42:2363--2373.

\bibitem{kongsgaard_adaptive_2017}
Kongsgaard M, {et~al.}
\newblock (2017) Adaptive immune responses to booster vaccination against
  yellow fever virus are much reduced compared to those after primary
  vaccination.
\newblock \emph{Scientific Reports} 7:662.

\bibitem{james_yellow_2013}
James EA, {et~al.}
\newblock (2013) Yellow fever vaccination elicits broad functional {CD}4+ {T}
  cell responses that recognize structural and nonstructural proteins.
\newblock \emph{Journal of Virology} 87:12794--12804.

\bibitem{dewitt_dynamics_2015}
DeWitt WS, {et~al.}
\newblock (2015) Dynamics of the cytotoxic {T} cell response to a model of
  acute viral infection.
\newblock \emph{Journal of Virology} 89:4517--4526.

\bibitem{pogorelyy_precise_2018}
Pogorelyy MV, {et~al.}
\newblock (2018) Precise tracking of vaccine-responding {T} cell clones reveals
  convergent and personalized response in identical twins.
\newblock \emph{Proceedings of the National Academy of Sciences of the United
  States of America} 115:12704--12709.

\bibitem{akondy_origin_2017}
Akondy RS, {et~al.}
\newblock (2017) Origin and differentiation of human memory {CD}8 {T} cells
  after vaccination.
\newblock \emph{Nature} 552:362--367.

\bibitem{de_melo_t-cell_2013}
de~Melo AB, {et~al.}
\newblock (2013) T-cell memory responses elicited by yellow fever vaccine are
  targeted to overlapping epitopes containing multiple {HLA}-{I} and -{II}
  binding motifs.
\newblock \emph{PLoS neglected tropical diseases} 7:e1938.

\bibitem{co_human_2002}
Co MDT, Terajima M, Cruz J, Ennis FA, Rothman AL
\newblock (2002) Human cytotoxic {T} lymphocyte responses to live attenuated
  17d yellow fever vaccine: identification of {HLA}-{B}35-restricted {CTL}
  epitopes on nonstructural proteins {NS}1, {NS}2b, {NS}3, and the structural
  protein {E}.
\newblock \emph{Virology} 293:151--163.

\bibitem{fuertes_marraco_long-lasting_2015}
Fuertes~Marraco SA, {et~al.}
\newblock (2015) Long-lasting stem cell-like memory {CD}8+ {T} cells with a
  naïve-like profile upon yellow fever vaccination.
\newblock \emph{Science Translational Medicine} 7:282ra48.

\bibitem{wieten_single_2016}
Wieten RW, {et~al.}
\newblock (2016) A {Single} 17d {Yellow} {Fever} {Vaccination} {Provides}
  {Lifelong} {Immunity}; {Characterization} of {Yellow}-{Fever}-{Specific}
  {Neutralizing} {Antibody} and {T}-{Cell} {Responses} after {Vaccination}.
\newblock \emph{PloS One} 11:e0149871.

\bibitem{dash_quantifiable_2017}
Dash P, {et~al.}
\newblock (2017) Quantifiable predictive features define epitope-specific {T}
  cell receptor repertoires.
\newblock \emph{Nature} 547:89--93.

\bibitem{glanville_identifying_2017}
Glanville J, {et~al.}
\newblock (2017) Identifying specificity groups in the {T} cell receptor
  repertoire.
\newblock \emph{Nature} 547:94--98.

\bibitem{schritt_repertoire_2019}
Schritt D, {et~al.}
\newblock (2019) Repertoire {Builder}: high-throughput structural modeling of
  {B} and {T} cell receptors.
\newblock \emph{Molecular Systems Design \& Engineering} 4:761--768.

\bibitem{pierce_flexible_2013}
Pierce BG, Weng Z
\newblock (2013) A flexible docking approach for prediction of {T} cell
  receptor-peptide-{MHC} complexes.
\newblock \emph{Protein Science: A Publication of the Protein Society}
  22:35--46.

\bibitem{appay_phenotype_2008}
Appay V, van Lier RAW, Sallusto F, Roederer M
\newblock (2008) Phenotype and function of human {T} lymphocyte subsets:
  {Consensus} and issues: {Phenotype} and {Function} of {Human} {T}
  {Lymphocyte} {Subsets}: {Consensus} and {Issues}.
\newblock \emph{Cytometry Part A} 73A:975--983.

\bibitem{sathaliyawala_distribution_2013}
Sathaliyawala T, {et~al.}
\newblock (2013) Distribution and compartmentalization of human circulating and
  tissue-resident memory {T} cell subsets.
\newblock \emph{Immunity} 38:187--197.

\bibitem{thome_spatial_2014}
Thome JJC, {et~al.}
\newblock (2014) Spatial map of human {T} cell compartmentalization and
  maintenance over decades of life.
\newblock \emph{Cell} 159:814--828.

\bibitem{bovay_t_2018}
Bovay A, {et~al.}
\newblock (2018) T cell receptor alpha variable 12-2 bias in the immunodominant
  response to {Yellow} fever virus.
\newblock \emph{European Journal of Immunology} 48:258--272.

\bibitem{zhang_high-throughput_2018}
Zhang SQ, {et~al.}
\newblock (2018) High-throughput determination of the antigen specificities of
  {T} cell receptors in single cells.
\newblock \emph{Nature Biotechnology}.

\bibitem{lyskov_rosettadock_2008}
Lyskov S, Gray JJ
\newblock (2008) The {RosettaDock} server for local protein-protein docking.
\newblock \emph{Nucleic Acids Research} 36:W233--238.

\bibitem{stuart_comprehensive_2019}
Stuart T, {et~al.}
\newblock (2019) Comprehensive {Integration} of {Single}-{Cell} {Data}.
\newblock \emph{Cell} 177:1888--1902.e21.

\bibitem{butler_integrating_2018}
Butler A, Hoffman P, Smibert P, Papalexi E, Satija R
\newblock (2018) Integrating single-cell transcriptomic data across different
  conditions, technologies, and species.
\newblock \emph{Nature Biotechnology} 36:411--420.

\bibitem{finak_mast:_2015}
Finak G, {et~al.}
\newblock (2015) {MAST}: a flexible statistical framework for assessing
  transcriptional changes and characterizing heterogeneity in single-cell {RNA}
  sequencing data.
\newblock \emph{Genome Biology} 16:278.

\bibitem{jeannet_essential_2010}
Jeannet G, {et~al.}
\newblock (2010) Essential role of the {Wnt} pathway effector {Tcf}-1 for the
  establishment of functional {CD}8 {T} cell memory.
\newblock \emph{Proceedings of the National Academy of Sciences of the United
  States of America} 107:9777--9782.

\bibitem{zhou_differentiation_2010}
Zhou X, {et~al.}
\newblock (2010) Differentiation and persistence of memory {CD}8(+) {T} cells
  depend on {T} cell factor 1.
\newblock \emph{Immunity} 33:229--240.

\bibitem{kaech_selective_2003}
Kaech SM, {et~al.}
\newblock (2003) Selective expression of the interleukin 7 receptor identifies
  effector {CD}8 {T} cells that give rise to long-lived memory cells.
\newblock \emph{Nature Immunology} 4:1191--1198.

\bibitem{jung_ccr7_2016}
Jung YW, Kim HG, Perry CJ, Kaech SM
\newblock (2016) {CCR}7 expression alters memory {CD}8 {T}-cell homeostasis by
  regulating occupancy in {IL}-7- and {IL}-15-dependent niches.
\newblock \emph{Proceedings of the National Academy of Sciences of the United
  States of America} 113:8278--8283.

\bibitem{schluns_interleukin-7_2000}
Schluns KS, Kieper WC, Jameson SC, Lefrançois L
\newblock (2000) Interleukin-7 mediates the homeostasis of naïve and memory
  {CD}8 {T} cells in vivo.
\newblock \emph{Nature Immunology} 1:426--432.

\bibitem{patil_precursors_2018}
Patil VS, {et~al.}
\newblock (2018) Precursors of human {CD}4+ cytotoxic {T} lymphocytes
  identified by single-cell transcriptome analysis.
\newblock \emph{Science Immunology} 3.

\bibitem{harari_distinct_2009}
Harari A, Bellutti~Enders F, Cellerai C, Bart PA, Pantaleo G
\newblock (2009) Distinct profiles of cytotoxic granules in memory {CD}8 {T}
  cells correlate with function, differentiation stage, and antigen exposure.
\newblock \emph{Journal of Virology} 83:2862--2871.

\bibitem{bratke_differential_2005}
Bratke K, Kuepper M, Bade B, Virchow JC, Luttmann W
\newblock (2005) Differential expression of human granzymes {A}, {B}, and {K}
  in natural killer cells and during {CD}8+ {T} cell differentiation in
  peripheral blood.
\newblock \emph{European Journal of Immunology} 35:2608--2616.

\bibitem{truong_killer-like_2019}
Truong KL, {et~al.}
\newblock (2019) Killer-like receptors and {GPR}56 progressive expression
  defines cytokine production of human {CD}4+ memory {T} cells.
\newblock \emph{Nature Communications} 10:2263.

\bibitem{takata_three_2006}
Takata H, Takiguchi M
\newblock (2006) Three memory subsets of human {CD}8+ {T} cells differently
  expressing three cytolytic effector molecules.
\newblock \emph{Journal of Immunology (Baltimore, Md.: 1950)} 177:4330--4340.

\bibitem{araki_translation_2017}
Araki K, {et~al.}
\newblock (2017) Translation is actively regulated during the differentiation
  of {CD}8+ effector {T} cells.
\newblock \emph{Nature Immunology} 18:1046--1057.

\bibitem{chu_longitudinal_2019}
Chu ND, {et~al.}
\newblock (2019) Longitudinal immunosequencing in healthy people reveals
  persistent {T} cell receptors rich in highly public receptors.
\newblock \emph{BMC immunology} 20:19.

\bibitem{monath_clinical_2002}
Monath TP, {et~al.}
\newblock (2002) Clinical proof of principle for {ChimeriVax}: recombinant
  live, attenuated vaccines against flavivirus infections.
\newblock \emph{Vaccine} 20:1004--1018.

\bibitem{moore_dependence_2019}
Moore JR, {et~al.}
\newblock (2019) Dependence of {CD}8 {T} {Cell} {Response} upon {Antigen}
  {Load} {During} {Primary} {Infection} : {Analysis} of {Data} from {Yellow}
  {Fever} {Vaccination}.
\newblock \emph{Bulletin of Mathematical Biology} 81:2553--2568.

\bibitem{reinhardt_development_1998}
Reinhardt B, Jaspert R, Niedrig M, Kostner C, L'age-Stehr J
\newblock (1998) Development of viremia and humoral and cellular parameters of
  immune activation after vaccination with yellow fever virus strain 17d: a
  model of human flavivirus infection.
\newblock \emph{Journal of Medical Virology} 56:159--167.

\bibitem{pogorelyy_persisting_2017}
Pogorelyy MV, {et~al.}
\newblock (2017) Persisting fetal clonotypes influence the structure and
  overlap of adult human {T} cell receptor repertoires.
\newblock \emph{PLoS computational biology} 13:e1005572.

\bibitem{shugay_towards_2014}
Shugay M, {et~al.}
\newblock (2014) Towards error-free profiling of immune repertoires.
\newblock \emph{Nature Methods} 11:653--655.

\bibitem{bolotin_mixcr:_2015}
Bolotin DA, {et~al.}
\newblock (2015) {MiXCR}: software for comprehensive adaptive immunity
  profiling.
\newblock \emph{Nature Methods} 12:380--381.

\bibitem{robinson_edger:_2010}
Robinson MD, McCarthy DJ, Smyth GK
\newblock (2010) {edgeR}: a {Bioconductor} package for differential expression
  analysis of digital gene expression data.
\newblock \emph{Bioinformatics (Oxford, England)} 26:139--140.

\bibitem{grant_bio3d:_2006}
Grant BJ, Rodrigues APC, ElSawy KM, McCammon JA, Caves LSD
\newblock (2006) Bio3d: an {R} package for the comparative analysis of protein
  structures.
\newblock \emph{Bioinformatics (Oxford, England)} 22:2695--2696.

\bibitem{borrman_atlas:_2017}
Borrman T, {et~al.}
\newblock (2017) {ATLAS}: {A} database linking binding affinities with
  structures for wild-type and mutant {TCR}-{pMHC} complexes.
\newblock \emph{Proteins} 85:908--916.

\end{thebibliography}

\setcounter{figure}{0}
\setcounter{table}{0}
\renewcommand{\thefigure}{S\arabic{figure}}
\renewcommand{\thetable}{S\arabic{table}}

\begin{figure*}[p]
\noindent\includegraphics[width=0.75\linewidth]{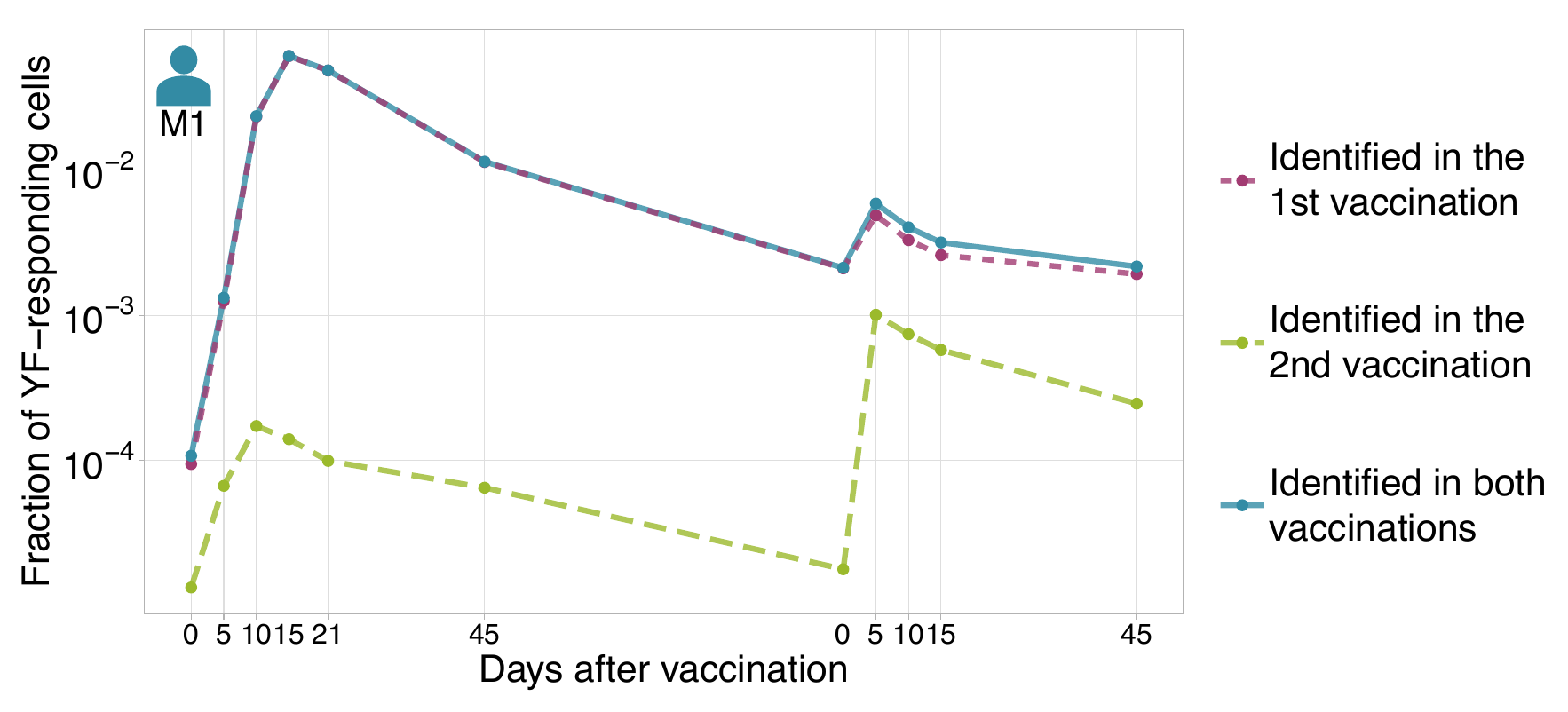}
\caption{{\ch The fraction of YF-responding cells as a proportion of all T-cells, measured by cumulative frequency of YF-responding TCR beta clonotypes of donor M1 identified by edgeR using timepoints after first vaccination (dashed purple), or after the second vaccination (dashed green). Solid blue line shows the sum of purple and green curves (clonotypes identified as expanded after first or after second immunization). }
}
\end{figure*}
\begin{figure*}[p]
\noindent\includegraphics[width=0.4\linewidth]{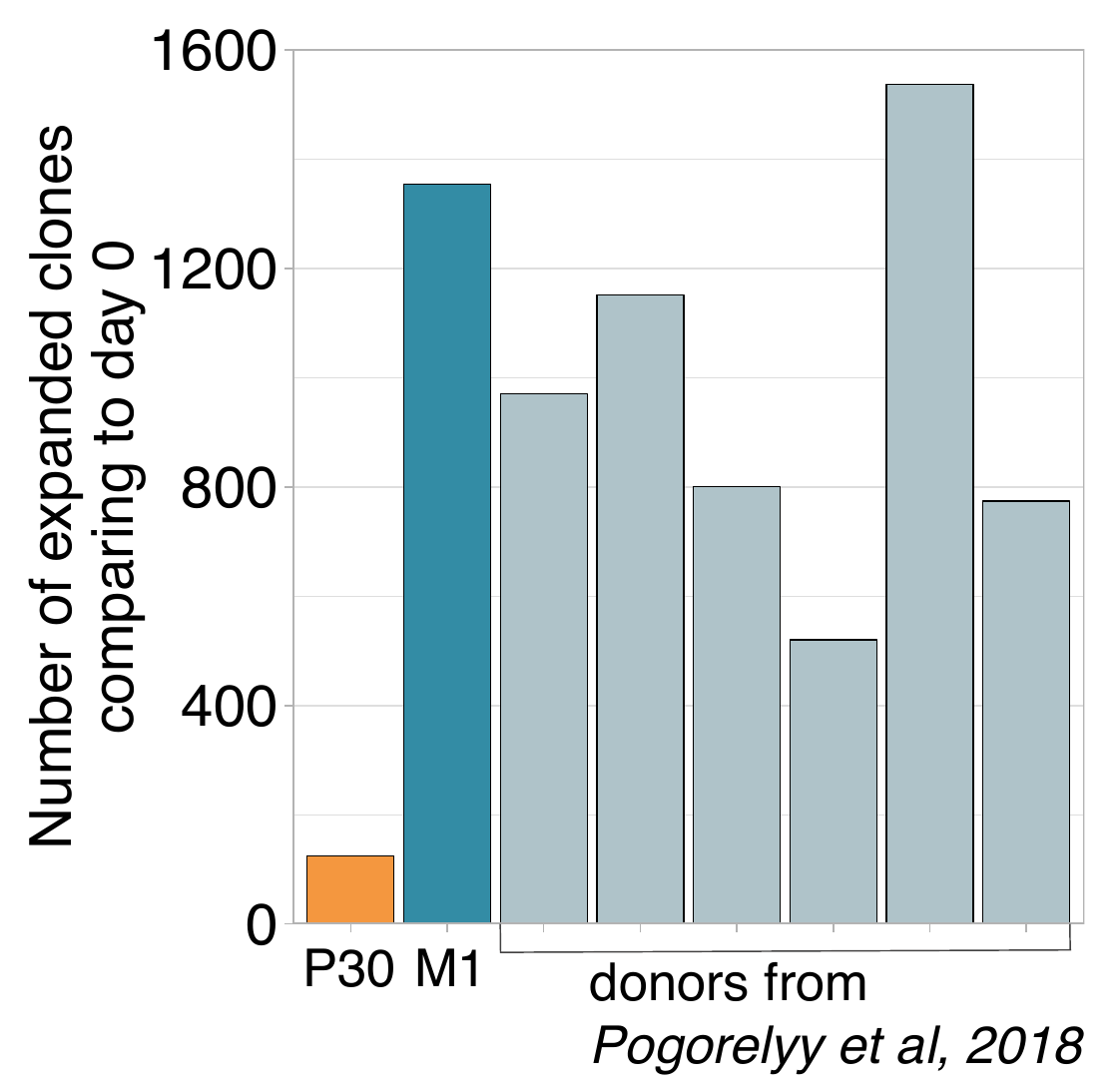}
\caption{{\bf Number of significantly expanded TCR beta clonotypes between day 0 and 15 identified by the edgeR software.} The donor revaccinated 30 years after the primary immunization has significantly fewer expanded clonotypes than any primary vaccinee.}
\end{figure*}
\begin{figure*}[p]
\noindent\includegraphics[width=\linewidth]{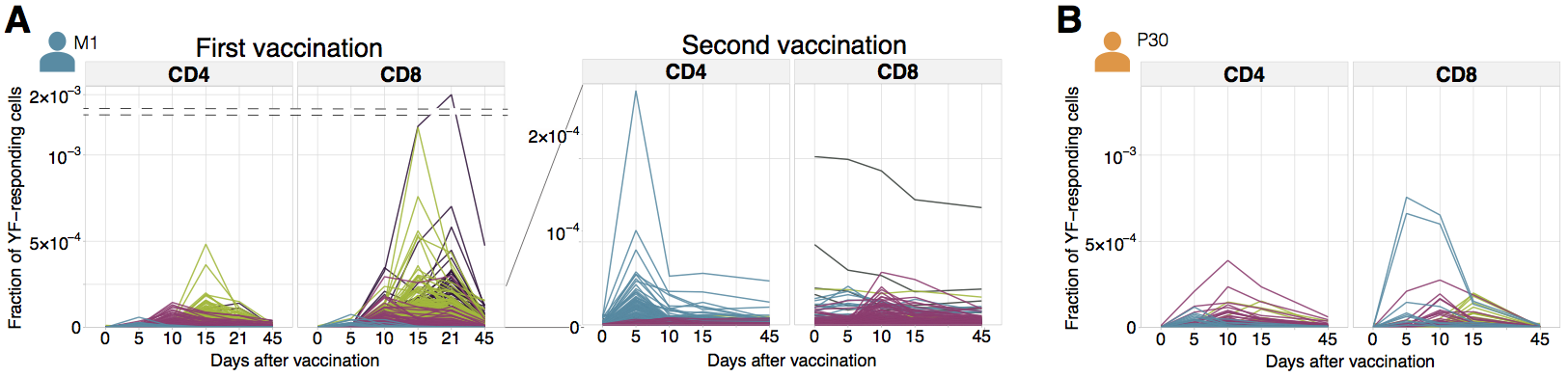}
\caption{{\bf Individual clonal trajectories of all YF-responding clonotypes.
} Frequency of each YF-responding clonotype {\ch in bulk TCR repertoire} as a function of time. Individual clones show remarkable expansion after the primary response ({\bf A}, left panel) and secondary response both 18 months ({\bf A}, right panel) and 30 years ({\bf B.}) after the primary vaccination. Color indicates the time of the response peak for each clonotype: blue for a peak at day 5, pink at day 10, green at day 15 and purple at day 21.}
\end{figure*}
\begin{figure*}[p]
\noindent\includegraphics[width=0.4\linewidth]{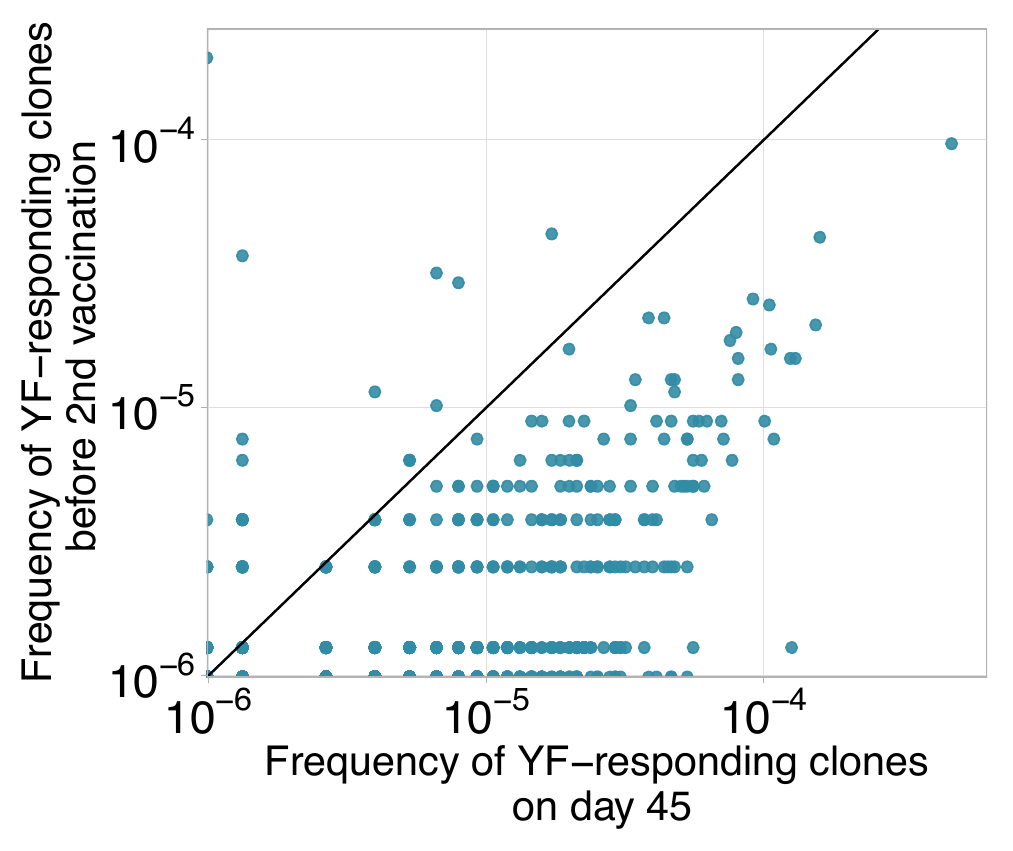}
\caption{{\bf Decay of YF-responding clonotypes between primary and secondary immunization.} Frequencies of YF-responding clones on day 45 of the primary immunization of donor M1 versus their frequencies 18 months later, before the second immunization). Diagonal line shows identity.}
\end{figure*}
\begin{figure*}[p]
\noindent\includegraphics[width=0.75\linewidth]{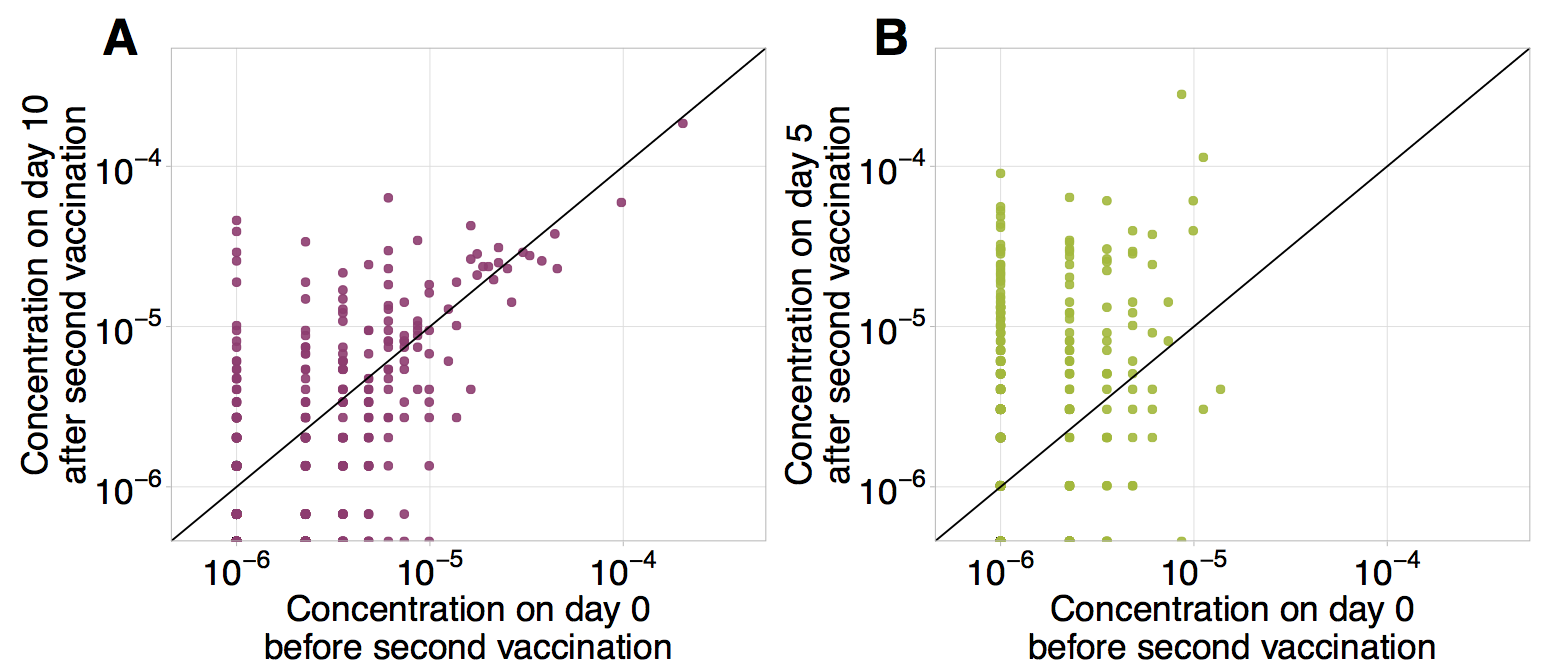}
\caption{{\bf Frequencies of  CD8+ (A) and CD4+ (B) clonotypes having responded to the primary YFV17D immunization in bulk before (x-axis) versus at the peak of the response to booster immunization (y-axis).} Diagonal line shows identity.}
\end{figure*}
\begin{figure*}[p]
\noindent\includegraphics[width=\linewidth]{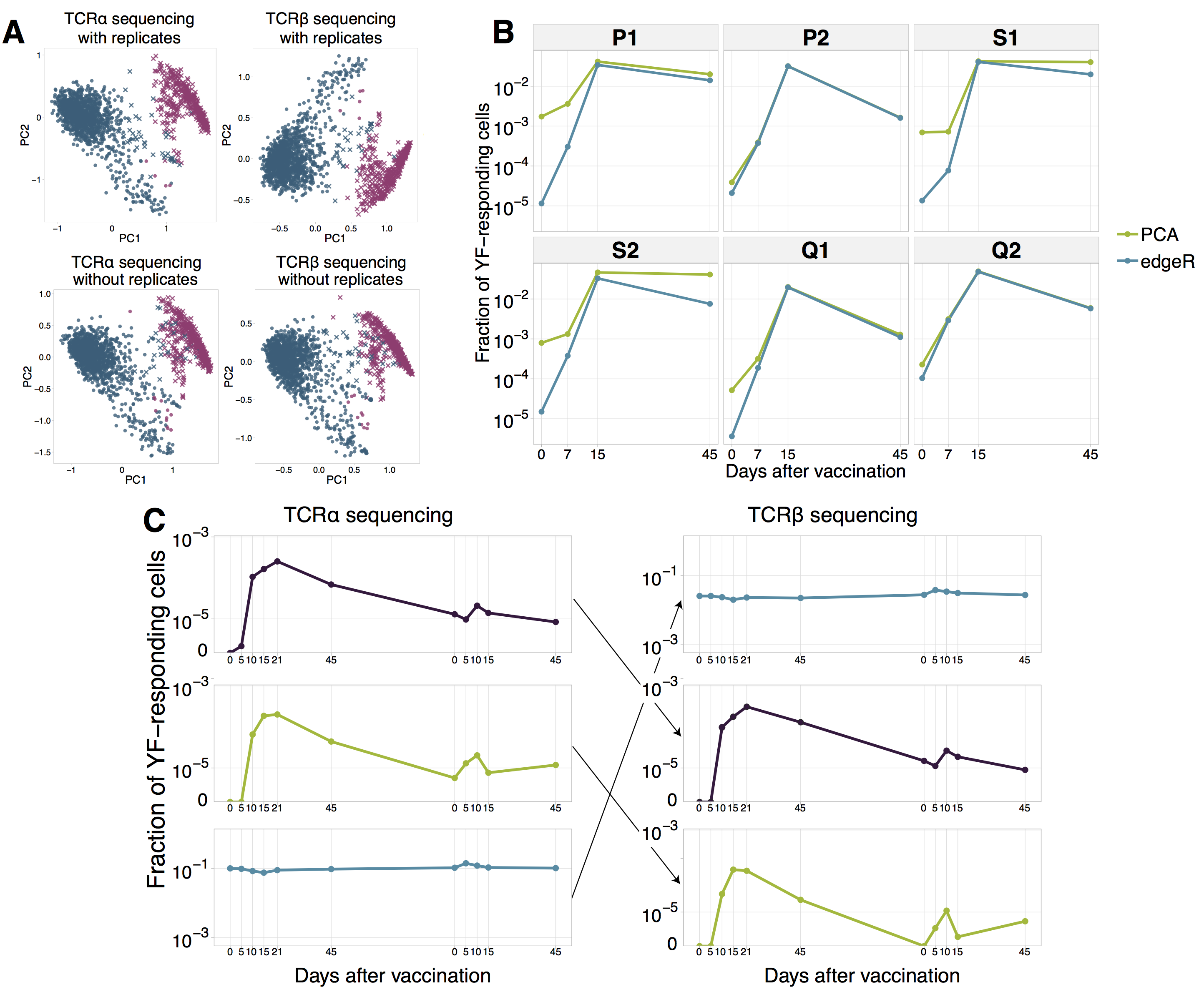}
\caption{\ch {\bf A.}{\bf YF-responding clones identified using hierarchical clustering of clonal time traces with and without biological replicates.} The plot shows two first  principal components  (x and y-axis) of the matrix, where rows are clonotypes and columns are normalized frequencies of these clonotypes on timepoints before and after primary immunization of donor M1. The frequency of each clonotype was normalized by its peak concentration. Pink color shows expanded clonotypes identified with edgeR. Two clusters (circles and crosses) were identified using hierarchical clustering. Similar results were obtained for both TCR alpha (left column) and TCR beta (right column) sequencing, with (top row) and without (bottom row) biological replicates for every timepoint.{\bf B.}{\bf Dynamics of YF-responding clonotypes after primary vaccination (data from Pogorelyy et al. 2018).} The cumulative frequency of YF-responding clonotypes defined as significantly expanded by edgeR is shown in blue. The green line indicates the cumulative frequency of responding clonotypes identified by hierarchical clustering of individual clonal trajectories. For the clustering procedure, only frequencies of biological replicate 1 of the bulk repertoire were used.{\bf C.} Examples of time traces for two YF-responding (purple and green) and one non-responding (blue) clonotypes in the TCR alpha repertoire (left), and their associated chain in TCR beta repertoire (right). The similarity of the alpha and beta traces of the same clone allows for computational alpha-beta pairing prediction. }
\end{figure*}
\begin{figure*}[p]
\noindent\includegraphics[width=\linewidth]{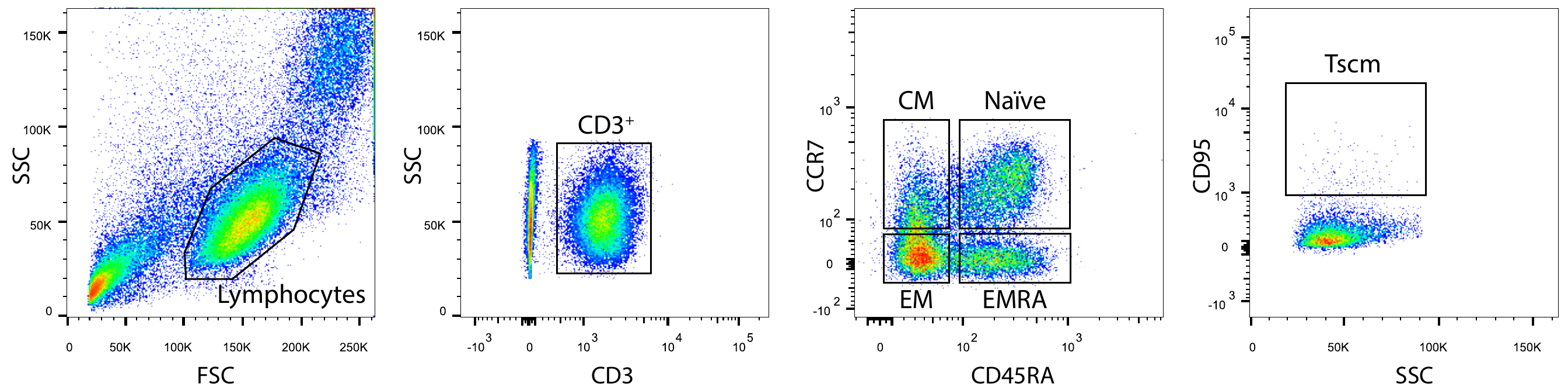}
\caption{{\bf Gating strategy for memory subpopulation.} Central memory (CM) cells were defined as CD3+CCR7+CD45RA-. Effector memory (EM) cells were CD3+CCR7-CD45RA-. Terminally differentiated effector memory (EMRA) cells were CD3+CCR7-CD45RA+. Stem-cell like memory (Tscm) cells were CD3+CCR7+CD45RA+CD95+.}
\end{figure*}
\begin{figure*}[p]
\noindent\includegraphics[width=0.75\linewidth]{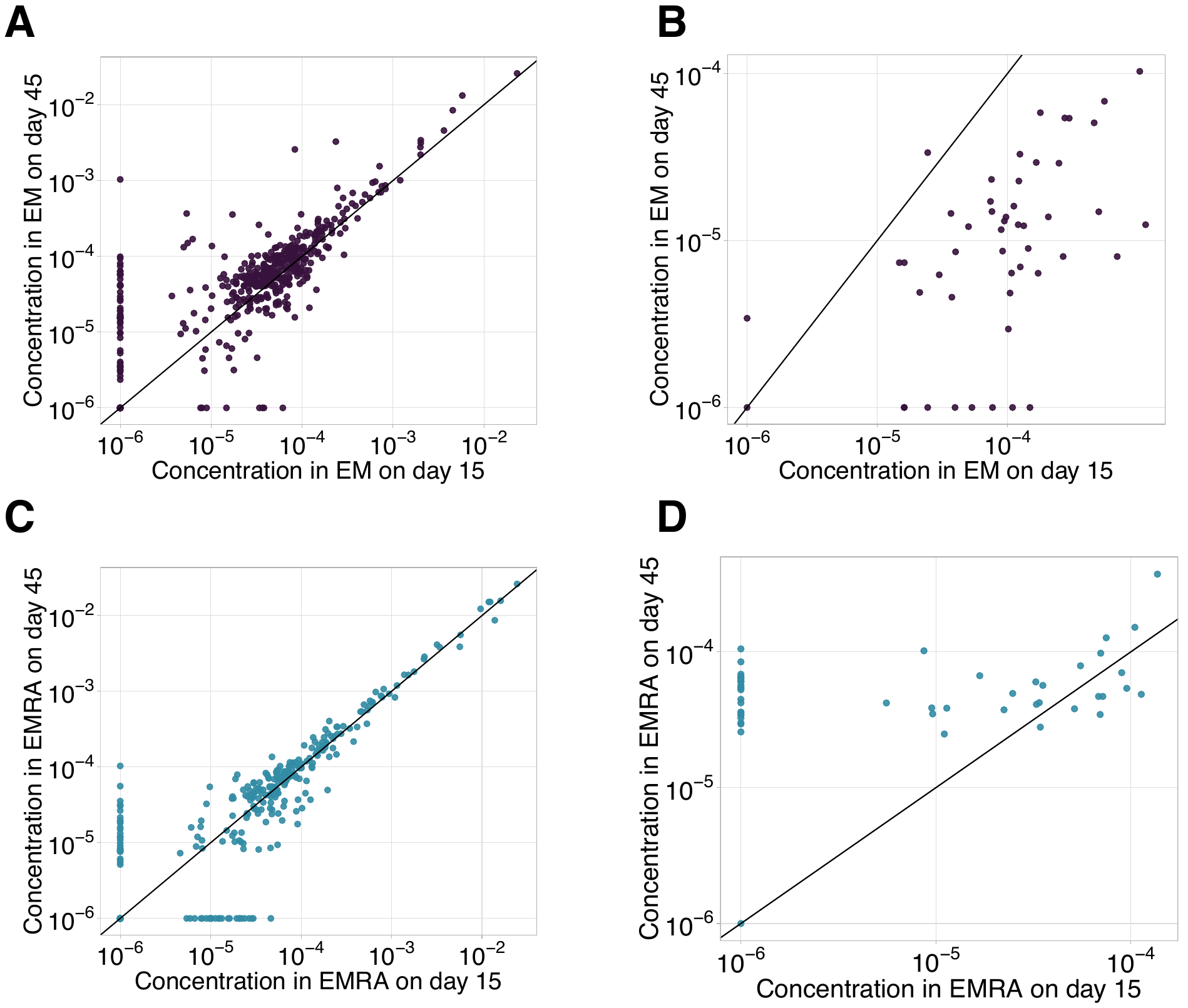}
\caption{{\bf EM-EMRA transition and decay {\ch of CD8+ clones} between day 15 and day 45.} We plot concentrations of EM (A, B) and EMRA (C, D) {\ch of CD8+} cells of each clone with $\geqslant 30$ UMI on day 45 in the bulk repertoire for non-YF-responding (A, C) and YF-responding (B, D) {\ch CD8+ clones} on day 15 (x-axis) versus day 45 (y-axis). Diagonal line shows identity.}
\end{figure*}
\begin{figure*}[p]
\noindent\includegraphics[width=0.75\linewidth]{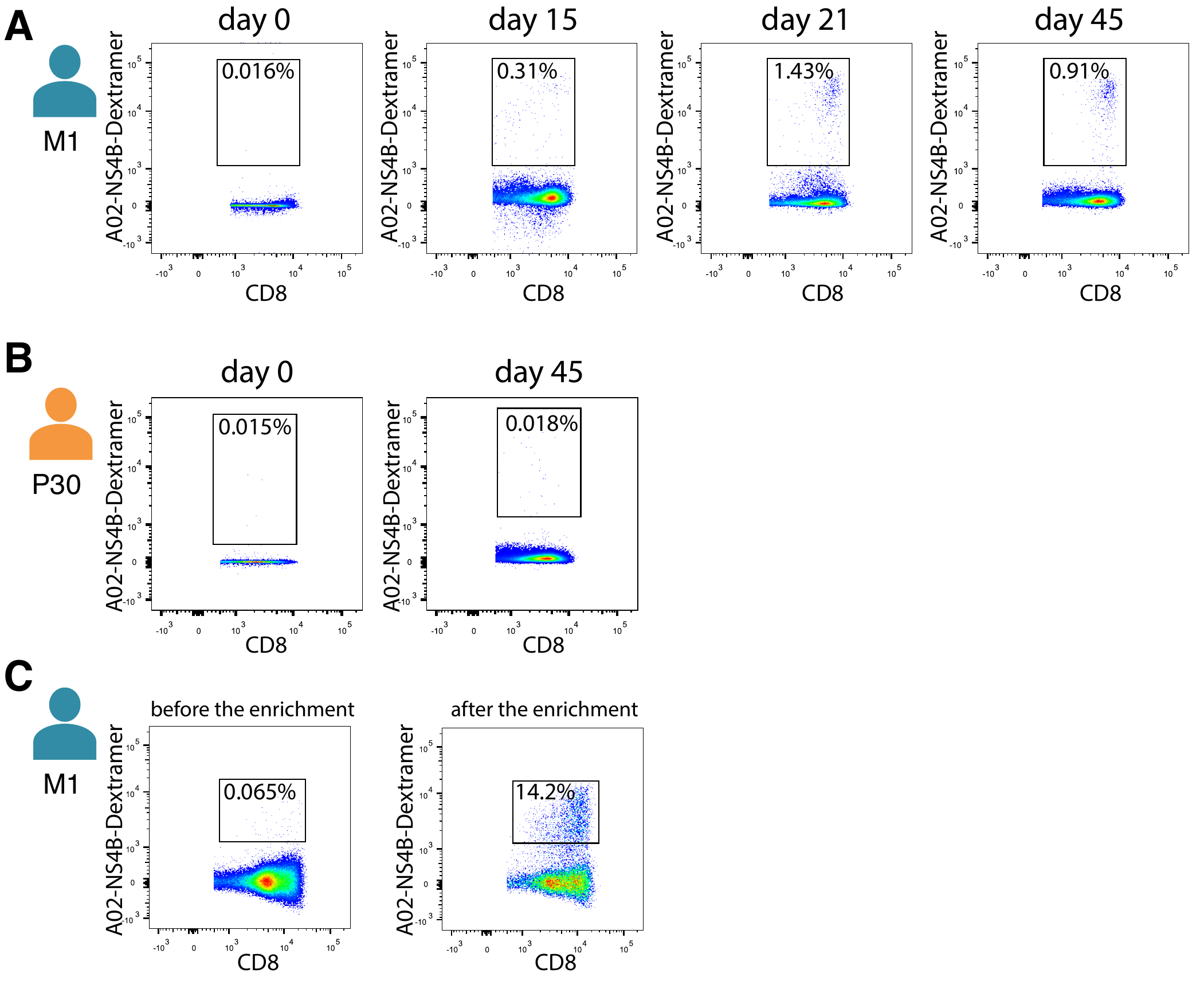}
\caption{{\bf Isolation of NS4B-specific T-cells of donor M1 (A) and donor P30 (B) on different timepoints after YF vaccination.}{\bf C. Number of NS4B-dextramer-positive cells before (left) and after (right) enrichment on the magnetic beads.} FACS was performed on donor M1 before the second immunization.}
\end{figure*}
\begin{figure*}[p]
\noindent\includegraphics[width=0.5\linewidth]{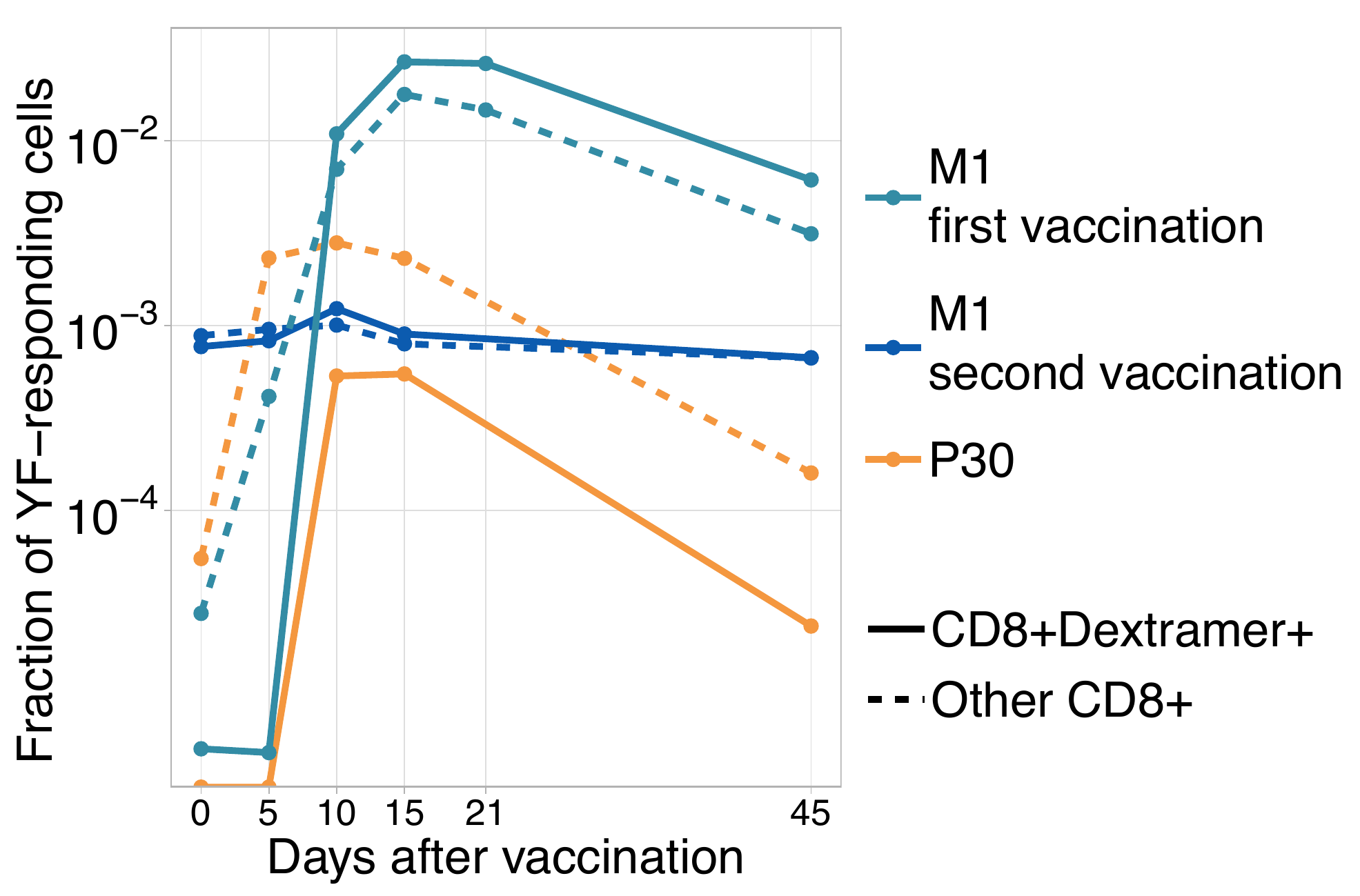}
\caption{{\bf  Dynamics of immunodominant response and other responses.} Total frequency of YF-responding NS4B-dextramer positive (solid line) and other YF-responding  CD8 clonotypes (dashed line) is plotted on different timepoints after immunization. All clonotypes are called YF-responding using edgeR. }
\end{figure*}
\begin{figure*}[p]
\noindent\includegraphics[width=\linewidth]{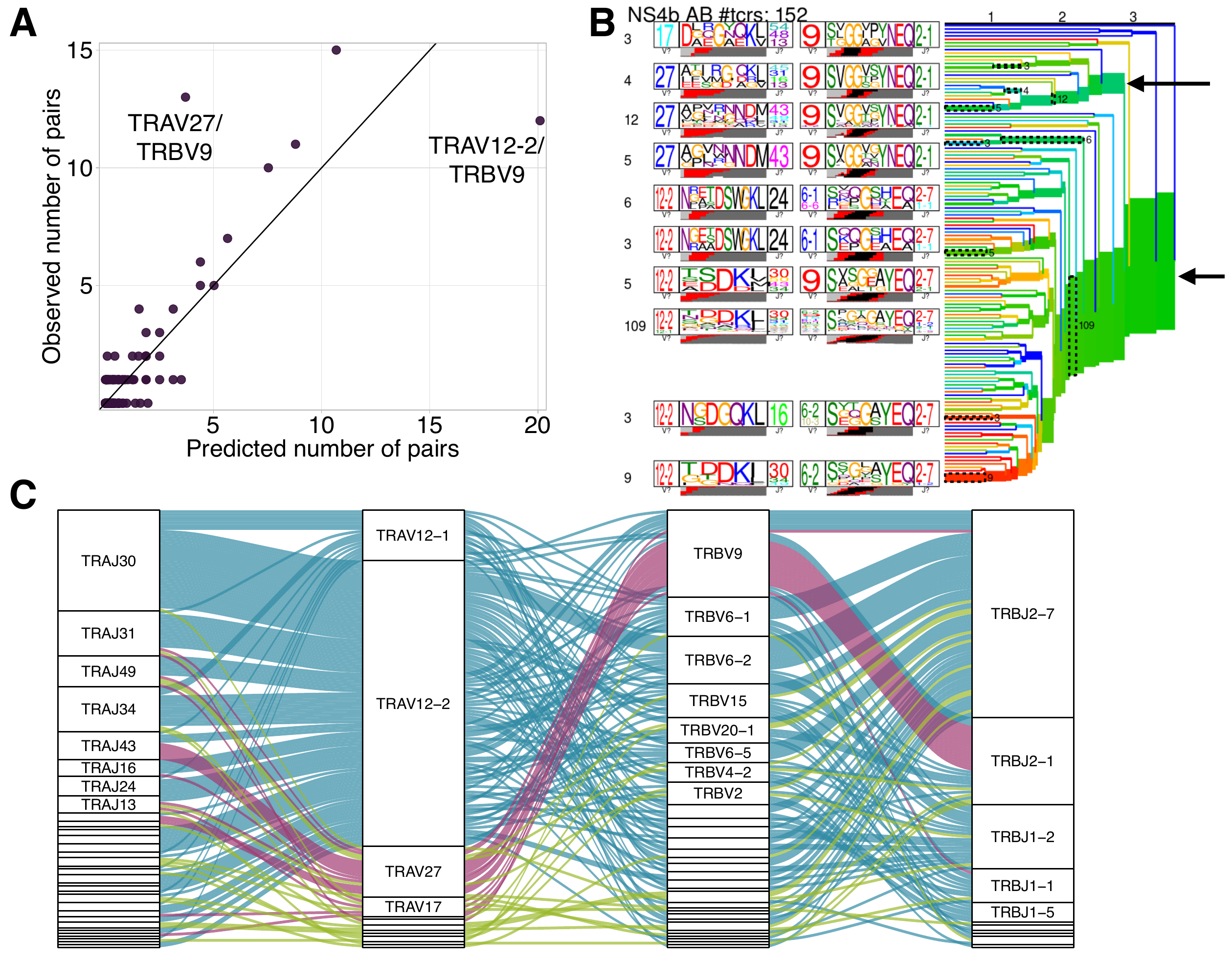}
\caption{{\bf A.}{\bf TRAV-TRBV pairing in NS4B-dextramer-positive TCRs.} Each dot shows TRAV-TRBV combination. The observed number of clonotypes using this combination in TCR is plotted against the number expected under random pairing from TRAV and TRBV frequencies. TRBV9 is expected to form more pairs with TRAV12-2 but pairs with TRAV27 instead, suggesting the existence of selective pressure on the choice of both chains.{\bf B.}{\bf Results of TCRdist hierarchical clustering of paired scTCR repertoire of NS4B-specific cells.} The two largest branches indicated with arrows correspond to TRAV12-2 and TRAV27-TRBV9 motifs.{\bf C.}{\bf Pairings of J-segments and V-segments of TCR alpha (left) to V-segments and J-segments of TCR beta (right) in scTCRseq of NS4B-specific T-cells.} The height of each box is proportional to the number of unique clones with a given gene segment. The width of the ribbons is proportional to the frequency of segment combinations. NS4B-specific TCRs have two main binding modes, defined by the TRAV12 segment family paired to almost any TRBV (blue) and by the TRAV27 segment paired preferentially with TRBV9 (pink). Other combinations are shown in green.}
\end{figure*}
\begin{figure*}[p]
\noindent\includegraphics[width=\linewidth]{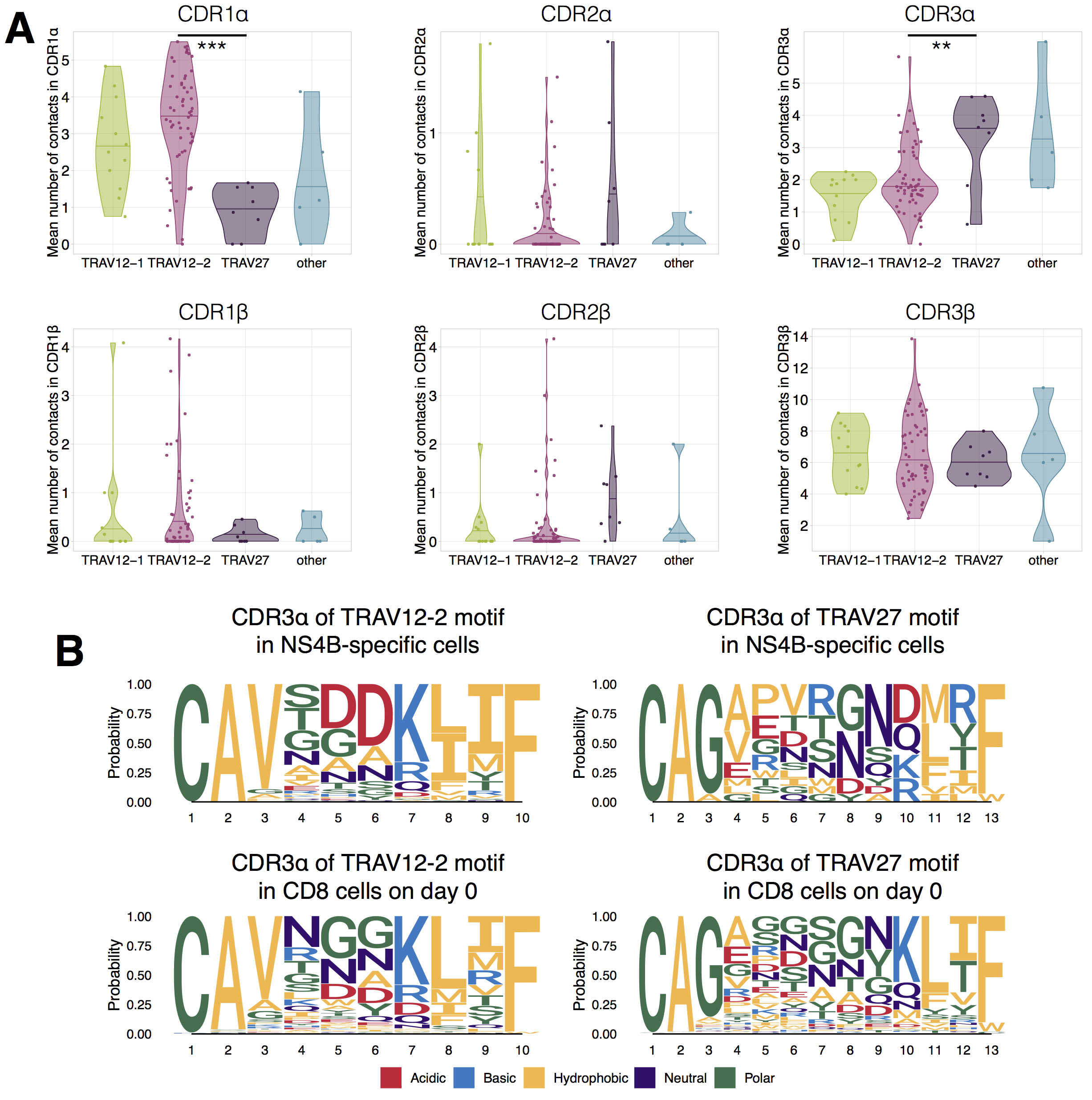}
\caption{{\bf A.}{\bf Average number of contacts to the LLWNGPMAV peptide in complementary determining regions of TCR alpha (top row) and TCR beta (bottom row) chains.} TCRs with TRAV12 segment (green and pink) have significantly more contacts (
Mann Whitney U-test p-value = 0.00015) in CDR1$\alpha$ than TCRs with TRAV27 (purple). On the other hand TCRs with TRAV27 have more contacts in CDR3$\alpha$ than TRAV12-2 TCRs (
Mann Whitney U-test p-value = 0.009). No significant difference in the number of contacts was observed for these binding modes in CDRs of the TCR beta chain. {\bf B.} {\bf Frequency of amino acids in CDR3s of clones with TRAV12-2 and TRAV27 V-segments in dextramer-sorted NS4B-specific clonotypes and bulk CD8 clonotypes prior to the vaccination.} For TRAV12-2 motif frequency distribution for TRAV12-2 is close to observed in bulk, suggesting absence of strong selection for certain amino acids in certain positions.}
\end{figure*}
\begin{figure*}[p]
\noindent\includegraphics[width=\linewidth]{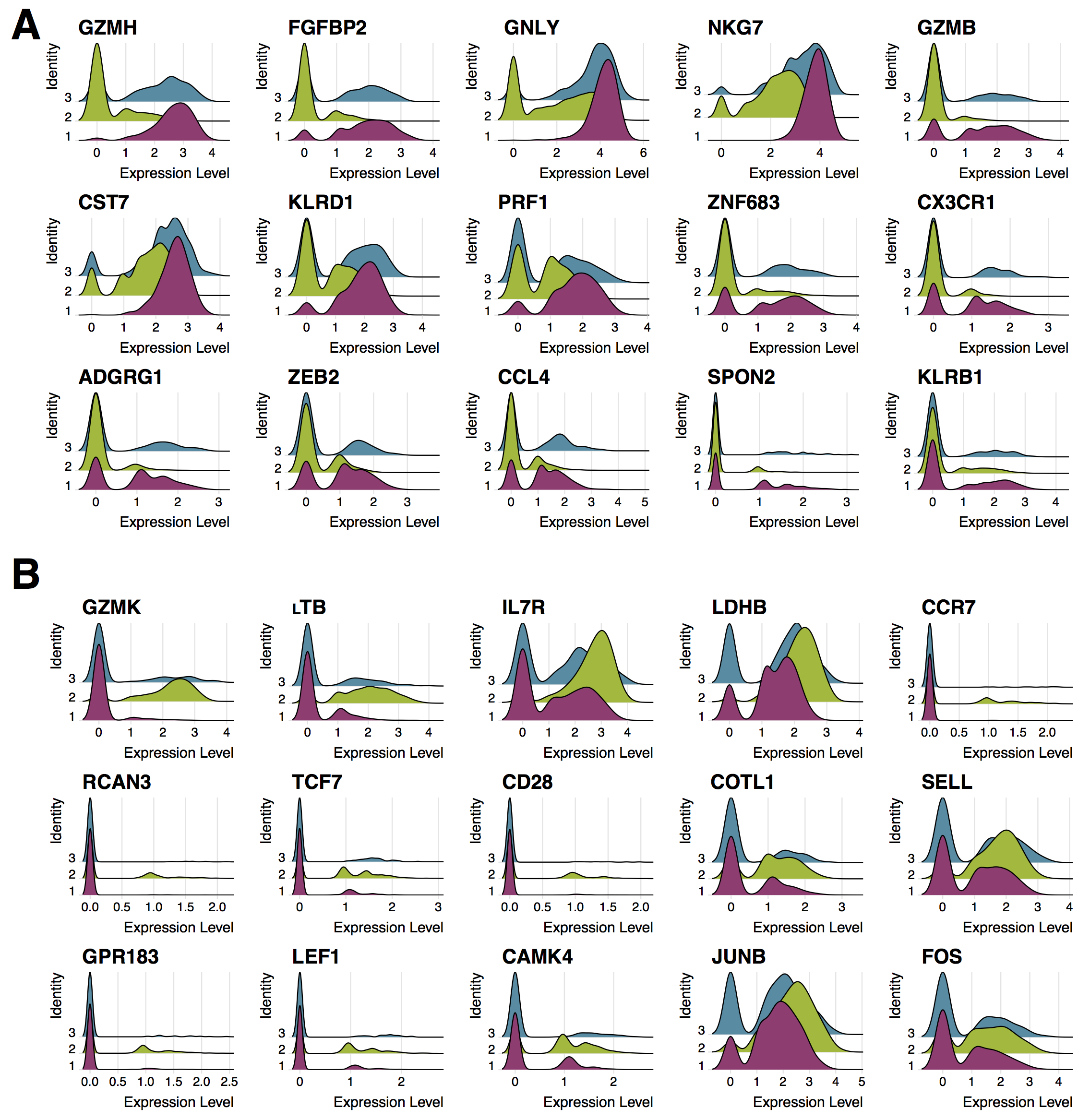}
\caption{{\bf A.}{\bf Expression of 15 genes most characteristic of cluster 1 in cells corresponding to clusters 1 (pink), 2 (green) and 3 (blue).} Cluster 3 has the intermediate phenotype.{\bf B.}{\bf Expression of 15 genes most characteristic of cluster 2 
in cells corresponding to clusters 1 (pink), 2 (green) and 3 (blue).} Cluster 3 has the intermediate phenotype.}
\end{figure*}
\begin{figure*}[p]
\noindent\includegraphics[width=0.75\linewidth]{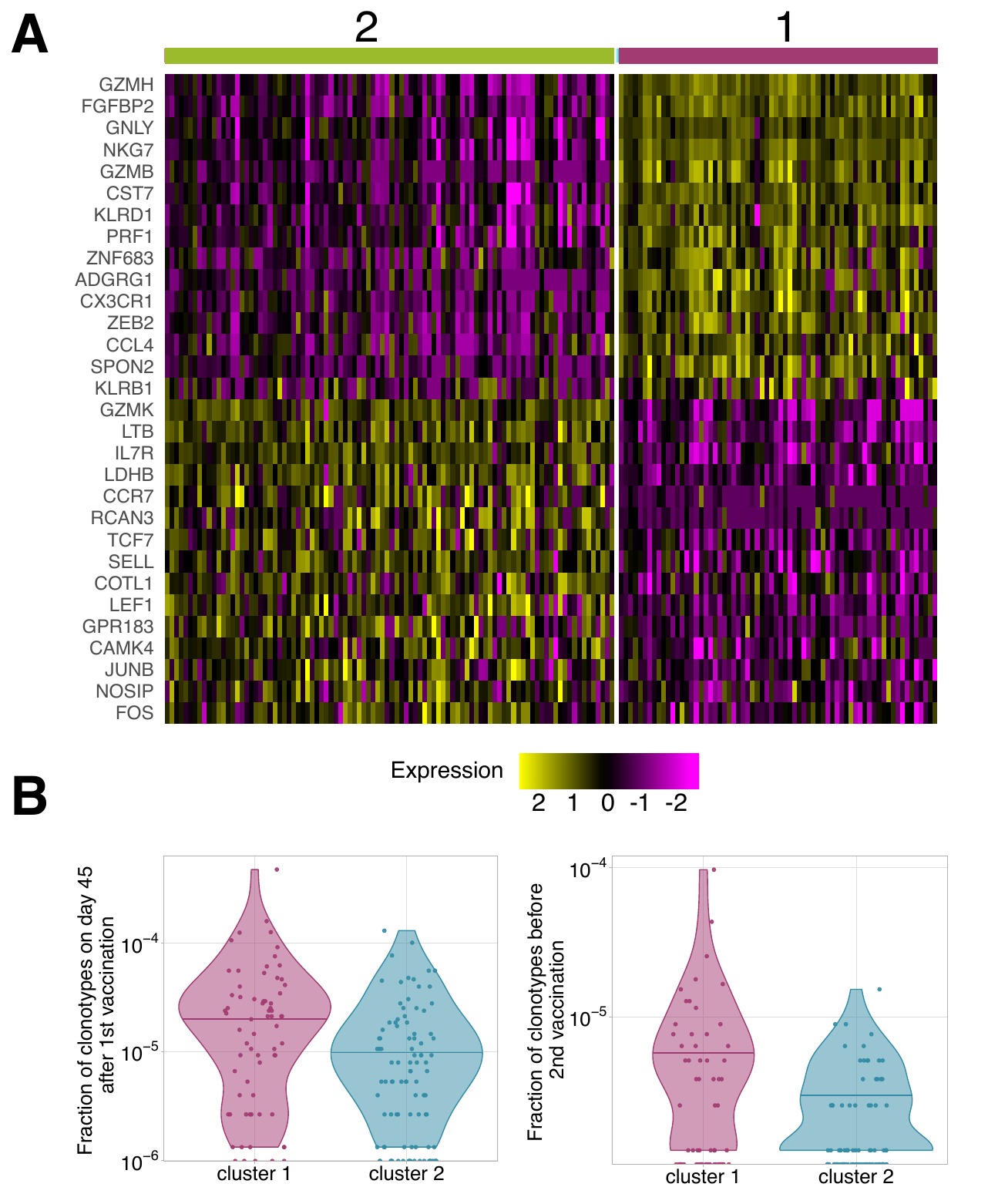}
\caption{{\bf A.}{\bf Genes differentially expressed between clonotypes.} Gene expression in each cell was averaged over the clonotypes before differential gene expression analysis. Unsupervised clustering shows 2 clusters with very similar gene expression to clusters 1 and 2 observed on scRNAseq of individual cells (Fig. 5B).{\bf B.}{\bf Frequency of clonotypes corresponding to cluster 1 and 2, after primary immunization (left), and 18 months later before the booster vaccination (right).} Clonotypes associated to cluster 1 are significantly more abundant on both these timepoints (Mann Whitney U-test A: p-value = 0.0003; B: p-value = 0.02447 ).}
\end{figure*}
\begin{figure*}[p]
\noindent\includegraphics[width=0.75\linewidth]{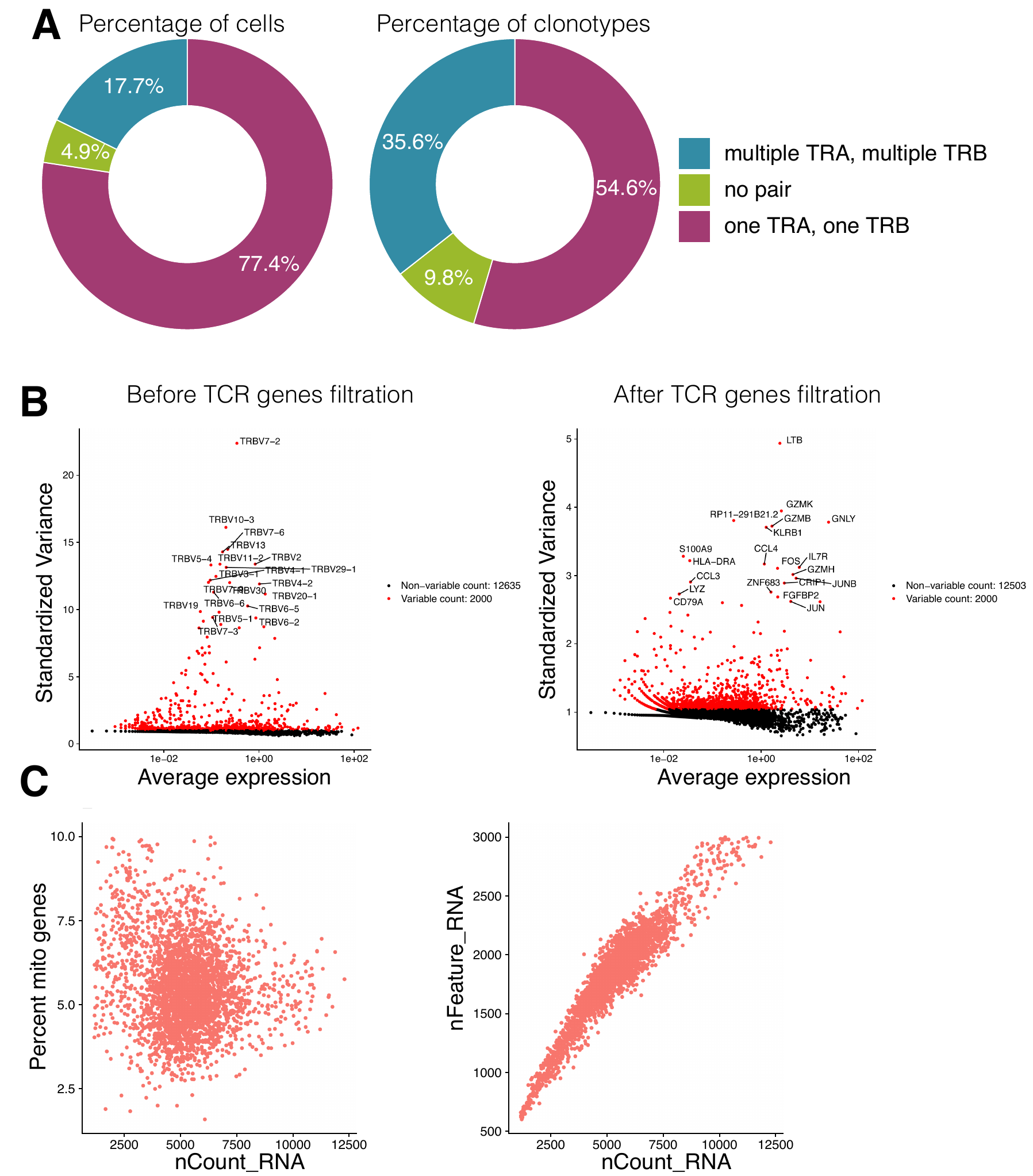}
\caption{{\bf A.}{\bf Proportion of cells (left) and clonotypes (right) in single-cell TCR sequencing data carrying different numbers of TCR alpha and TCR beta chains.}{\bf B.}{\bf Most variable genes in the dataset before ({\bf left}) and after ({\bf right}) the filtration of TCR related genes.}  TCR related genes were the source of unwanted variation in single-cell gene expression analysis and were removed from the data.{\bf C.}{\bf  Visualization of quality control  metrics in the single-cell gene expression experiment.} The relationship between the number of RNAs inside the cell (x-axis) and the percentage of mitochondrial genes (y-axis) is shown on the left. The relationship between the number of RNAs inside the cell (x-axis) and the number of genes (y-axis) is shown on the right. Cells that had more than 8\% of mitochondrial genes or more than 2700 total number of genes were discarded from further analysis.}
\end{figure*}

\FloatBarrier
\begin{table*}[p] \centering
\caption{HLA-typing results for donors M1 and P30}

\begin{tabular}{ccc}
\hline
Locus & M1 & P30  \\ \hline
A & 02:01:01/24:02:01 & 02:01:01/31:01:02 \\
B & 15:01:01/39:01:01 & 35:01:01/48:01:01  \\ 
C & 03:04:01/12:03:01 & 04:01:01/08:01:01  \\ 
DQB1 & 02:01:01/03:02:01 & 03:01:01/03:01:01  \\ 
DRB1 & 03:01:01/04:01:01 & 11:01:01/12:01:01  \\ 
DRB3 & 02:02:01 & 01:01:02/02:02:01 \\ 
DRB4 & 01:03:01 & -  \\ \hline
\end{tabular}
\end{table*}

\end{document}